\documentclass[a4paper,11pt]{article}

\usepackage{jheppub} 

\usepackage{amsmath,amssymb,array,calc,rotating,epsfig,psfrag, amscd, datetime}
\usepackage{verbatim}
\usepackage{xcolor}

\newcommand{\nc}{\newcommand}

\definecolor{cardinal}{rgb}{0.6,0,0}
\definecolor{darkgreen}{rgb}{0,0.5,0}
\definecolor{golden}{rgb}{0.92, 0.7, 0}
\definecolor{midnight}{rgb}{0, 0, 0.5}
\definecolor{darkblue}{rgb}{0.2, 0, 0.8}

\nc{\ra}{\rightarrow} 
\nc{\lra}{\leftrightarrow} 
\nc{\Ra}{\Rightarrow} 
\nc{\LRa}{\Leftightarrow} 
\nc{\blp}{{\big (}}
\nc{\brp}{{\big )}}
\nc{\Blp}{{\Big (}}
\nc{\Brp}{{\Big )}}
\nc{\bglp}{{\bigg (}}
\nc{\bgrp}{{\bigg )}}
\nc{\Bglp}{{\Bigg (}}
\nc{\Bgrp}{{\Bigg )}}
\nc{\slb}{{\rm [}}
\nc{\srb}{{\rm ]}}
\nc{\bslb}{{\rm \big [}}
\nc{\bsrb}{{\rm \big ]}}
\nc{\Bslb}{{\rm \Big [}}
\nc{\Bsrb}{{\rm \Big ]}}

\def\al{\alpha}

\def\eps{\epsilon}
\nc{\veps}{\varepsilon}
\def\gam{\gamma}

\def\lam{\lambda}
\def\om{\omega}

\nc{\vphi}{\varphi}
\def\tha{\theta}

\def\sig{\sigma}

\def\Gam{\Gamma}

\def\Lam{\Lambda}
\def\Om{\Omega}
\def\Sig{\Sigma}

\def\coeff#1#2{\relax{\textstyle {#1 \over #2}}\displaystyle}

\nc{\myvspace}{\rule[-1em]{0pt}{2.5em}}
\nc{\bea}{\begin{eqnarray}}
\nc{\eea}{\end{eqnarray}}
\nc{\be}{\begin{equation}}
\nc{\ee}{\end{equation}}
\nc{\barr}{\begin{array}}
\nc{\earr}{\end{array}}

\nc{\cA}{{\cal A}}
\nc{\cB}{ \cal B}

\def\cD{{\cal D}}

\nc{\cF}{{\cal F}}
\nc{\cG}{{\cal G}}

\nc{\cL}{{\cal L}}
\nc{\cM}{{\cal M}}
\def\N{{\cal N}}
\def\cN{{\cal N}}

\def\cO{{\cal O}}
\def\cP{{\cal P}}
\nc{\cQ}{{\cal Q}}
\nc{\cR}{{\cal R}}
\def\cS{{\cal S}}

\def\cV{{\cal V}}
\def\cV{{\cal V}}

\def\cZ{{\cal Z}}
\nc{\cQd}{\cQ^{\dagger}}
\nc{\cRd}{\cR^{\dagger}}
\nc{\BB}{{\mathbb B}}
\nc{\CC}{{\mathbb C}}
\nc{\DD}{{\mathbb D}}
\nc{\EE}{{\mathbb E}}
\nc{\FF}{{\mathbb F}}
\nc{\GG}{{\mathbb G}}
\nc{\HH}{{\mathbb H}}
\nc{\JJ}{{\mathbb J}}
\nc{\MM}{{\mathbb M}}
\nc{\RR}{{\mathbb R}}
\nc{\PP}{{\mathbb P}}
\nc{\QQ}{{\mathbb Q}}
\nc{\UU}{{\mathbb U}}
\nc{\ZZ}{{\mathbb Z}}
\nc{\calone}{{\mathbb 1}}

\nc{\half}{\coeff{1}{2}}
\nc{\quarter}{\coeff{1}{4}}
\nc{\del}{\partial}

\nc{\delbar}{\bar\partial}
\nc{\thalf}{\frac{t}{2}}
\nc{\Spin}{\operatorname{Spin}}
\nc{\SO}{\operatorname{SO}}

\nc{\Sp}{{\rm Sp}}
\nc{\com}[2]{{ \left[ #1, #2 \right] }}
\nc{\acom}[2]{{ \left\{ #1, #2 \right\} }}
\nc{\rr}{\rightarrow}
\nc{\p}{\partial}
\nc{\LT}{{\LL_\T}}
\nc{\Tr}{{\rm Tr}}
\nc{\tr}{{\rm tr}}
\nc{\Adag}{A^{\dagger}}
\nc{\AdagI}{A^{\dagger I}}
\nc{\AdagJ}{A^{\dagger J}}
\nc{\AdagK}{A^{\dagger K}}
\nc{\AdagL}{A^{\dagger L}}
\nc{\AdagM}{A^{\dagger M}}
\nc{\Bdag}{B^{\dagger}}
\nc{\BdagI}{B^{\dagger}_I}
\nc{\BdagJ}{B^{\dagger}_J}
\nc{\BdagK}{B^{\dagger}_K}
\nc{\BdagL}{B^{\dagger}_L}
\nc{\BdagM}{B^{\dagger}_M}
\nc{\Cdag}{C^{\dagger}}
\nc{\CdagI}{C^{\dagger I}}
\nc{\CdagJ}{C^{\dagger J}}
\nc{\CdagK}{C^{\dagger K}}
\nc{\Ddag}{D^{\dagger}}
\nc{\DdagI}{D^{\dagger I}}
\nc{\DdagJ}{D^{\dagger J}}
\nc{\DdagK}{D^{\dagger K}}
\nc{\ttha}{\tilde{\theta}}
\nc{\ttau}{\tilde{\tau}}
\nc{\tTha}{\tilde{\Theta}}
\nc{\tphi}{\tilde{\phi}}
\nc{\tsig}{\tilde{\sig}}
\nc{\tom}{\widetilde{\om}}
\nc{\tOm}{\widetilde{\Om}}
\nc{\tlam}{\widetilde{\lam}}
\nc{\tLam}{\tilde{\Lam}}
\nc{\tSig}{\widetilde{\Sig}}
\nc{\tPhi}{\tilde{\Phi}}
\nc{\tPhibar}{\ol{\tPhi}}
\nc{\tPi}{\widetilde{\Pi}}
\nc{\tpsi}{\widetilde{\psi}}
\nc{\tPsi}{\tilde{\Psi}}
\nc{\tgam}{\widetilde{\gam}}
\nc{\tGam}{\widetilde{\Gam}}
\nc{\tzeta}{\tilde{\zeta}}
\nc{\tZeta}{\tilde{\Zeta}}
\nc{\teta}{\widetilde{\eta}}
\nc{\teps}{\tilde{\eps}}
\nc{\tveps}{\tilde{\veps}}
\nc{\tEta}{\tilde{\Eta}}
\nc{\tchi}{\tilde{\chi}}
\nc{\tChi}{\tilde{\Chi}}
\nc{\txi}{\tilde{\xi}}
\nc{\tXi}{\widetilde{\Xi}}
\nc{\tnu}{\tilde{\nu}}
\nc{\tmu}{\tilde{\mu}}

\nc{\tb}{\tilde b}
\nc{\tc}{\tilde c}
\nc{\te}{\tilde e}
\nc{\tf}{\tilde f}
\nc{\tg}{\tilde g}
\nc{\ti}{\tilde i}
\nc{\tj}{\tilde j}
\nc{\tk}{\tilde k}
\nc{\tl}{\tilde l}
\nc{\tm}{\tilde m}
\nc{\tn}{\tilde n}
\nc{\tp}{\tilde{p}}
\nc{\tq}{\widetilde{q}}
\nc{\ts}{{\tilde s}}
\nc{\tu}{{\tilde u}}
\nc{\tv}{{\tilde v}}
\nc{\tw}{{\tilde w}}
\nc{\tx}{{\tilde x}}
\nc{\ty}{{\tilde y}}
\nc{\tz}{\tilde z}
\nc{\tA}{{\widetilde A}}
\nc{\tAbar}{{\ol \tA}}
\nc{\tB}{{\widetilde B}}
\nc{\tC}{{\widetilde C}}
\nc{\tD}{{\widetilde D}}
\nc{\tE}{{\widetilde E}}
\nc{\tF}{{\widetilde F}}
\nc{\tG}{{\widetilde G}}
\nc{\tH}{{\widetilde H}}
\nc{\tJ}{{\widetilde J}}
\nc{\tJbar}{{\ol {\tilde J}}}
\nc{\tK}{{\widetilde K}}
\nc{\tL}{{\widetilde L}}
\nc{\tcL}{{\widetilde \cL}}
\nc{\tM}{{\widetilde M}}
\nc{\tN}{{\widetilde N}}
\nc{\tcN}{{\widetilde \cN}}
\nc{\tP}{{\widetilde P}}
\nc{\tQ}{{\widetilde Q}}
\nc{\tR}{{\widetilde R}}
\nc{\tS}{\widetilde{S}}
\nc{\tT}{\widetilde{T}}
\nc{\tU}{\widetilde{U}}
\nc{\tV}{\widetilde{V}}
\nc{\tW}{\widetilde{W}}
\nc{\tcF}{\widetilde{{\cal F}}}
\nc{\tX}{\widetilde{X}}
\nc{\tY}{\widetilde{Y}}
\nc{\tcZ}{\tilde{\cZ}}
\nc{\tcZbar}{\ol{\tcZ}}

\nc{\ha}{\hat a}
\nc{\hb}{\hat b}
\nc{\hc}{\widehat c}
\nc{\hd}{\widehat d}
\nc{\he}{\widehat e}
\nc{\hf}{\widehat f}
\nc{\hg}{\widehat g}
\nc{\hh}{\widehat h}
\nc{\hm}{\widehat m}
\nc{\hn}{\widehat n}
\nc{\hp}{\widehat p}
\nc{\hr}{\widehat r}
\nc{\hs}{\widehat s}
\nc{\hv}{\widehat v}
\nc{\hw}{\widehat w}
\nc{\hx}{\widehat x}
\nc{\hy}{\widehat y}
\nc{\hz}{\widehat z}
\nc{\zhat}{\hat z}
\nc{\hA}{\widehat{A}}
\nc{\hB}{\widehat{B}}
\nc{\hC}{\widehat{C}}
\nc{\hD}{\widehat{D}}
\nc{\hE}{\widehat{E}}
\nc{\hF}{\widehat{F}}
\nc{\hcF}{\widehat{\cF}}
\nc{\hG}{\widehat{G}}
\nc{\hH}{\widehat{H}}
\nc{\hJ}{\widehat{J}}
\nc{\hK}{\widehat{K}}
\nc{\hL}{\widehat{L}}
\nc{\hcL}{\widehat{\cL}}
\nc{\hM}{\widehat M}
\nc{\hcM}{\widehat{\cM}}
\nc{\hN}{\widehat{N}}
\nc{\hO}{\widehat{O}}
\nc{\hP}{\widehat{P}}
\nc{\hQ}{\widehat{Q}}
\nc{\hcR}{\widehat{\cR}}
\nc{\hR}{\widehat{R}}
\nc{\hS}{\widehat{S}}
\nc{\hcS}{\widehat{\cS}}
\nc{\hT}{\widehat{T}}
\nc{\hU}{\widehat{U}}
\nc{\hV}{\widehat V}
\nc{\hcV}{\widehat \cV}
\nc{\hX}{\widehat X}

\nc{\heta}{\widehat{\eta}}
\nc{\hal}{\widehat \alpha}
\nc{\hphi}{\widehat{\phi}}
\nc{\hkap}{\hat{\kappa}}
\nc{\hchi}{\widehat{\chi}}
\nc{\hpsi}{\widehat{\psi}}
\nc{\hgam}{\widehat{\gam}}
\nc{\hPhi}{\hat{\Phi}}
\nc{\hPsi}{\hat{\Psi}}
\nc{\hGam}{\hat{\Gam}}
\nc{\omhat}{\widehat{\om}}
\nc{\htha}{\hat{\tha}}
\nc{\hrho}{\widehat{\rho}}
\nc{\hdel}{\widehat{\del}}

\nc{\w}{\wedge}

\nc{\vb}{\vec b}
\nc{\vc}{\vec c}
\nc{\vd}{\vec d}
\nc{\ve}{\vec e}
\nc{\vf}{\vec f}
\nc{\vg}{\vec g}
\nc{\vh}{\vec h}
\nc{\vp}{\vec p}
\nc{\vq}{\vec q}
\nc{\vr}{\vec r}
\nc{\vs}{\vec s}
\nc{\vv}{\vec v}
\nc{\vw}{\vec w}
\nc{\vx}{\vec x}
\nc{\vy}{\vec y}
\nc{\vz}{\vec z}

\nc{\vB}{\vec B}
\nc{\vC}{\vec C}
\nc{\vD}{\vec D}
\nc{\vE}{\vec E}
\nc{\vF}{\vec F}
\nc{\vG}{\vec G}
\nc{\vH}{\vec H}
\nc{\vP}{\vec P}
\nc{\vQ}{\vec Q}
\nc{\vR}{\vec R}
\nc{\vS}{\vec S}
\nc{\vV}{\vec V}
\nc{\vW}{\vec W}
\nc{\vX}{\vec X}
\nc{\vY}{\vec Y}
\nc{\vZ}{\vec Z}

\nc{\ol}{\overline}
\nc{\abar}{\ol{a}}
\nc{\bbar}{\ol{b}}
\nc{\cbar}{\ol{c}}
\nc{\dbar}{\ol{d}}
\nc{\ebar}{\ol{e}}
\nc{\fbar}{\ol{f}}
\nc{\ibar}{\ol{\imath}}
\nc{\jbar}{\ol{\jmath}}
\nc{\kbar}{\ol{k}}
\nc{\lbar}{\ol{l}}
\nc{\mbar}{\ol{m}}
\nc{\nbar}{\ol{n}}
\nc{\pbar}{\ol{p}}
\nc{\qbar}{\ol{q}}
\nc{\rbar}{\ol{r}}
\nc{\sbar}{\ol{s}}
\nc{\ubar}{\ol{u}}
\nc{\vbar}{\ol{v}}
\nc{\wbar}{\ol{w}}
\nc{\xbar}{\ol{x}}
\nc{\ybar}{\ol{y}}
\nc{\zbar}{\ol{z}}

\nc{\Abar}{\ol{A}}
\nc{\Bbar}{\ol{B}}
\nc{\Cbar}{\ol{C}}
\nc{\Dbar}{\ol{D}}
\nc{\Ebar}{\ol{E}}
\nc{\Fbar}{\ol{F}}
\nc{\Jbar}{\ol{J}}
\nc{\Kbar}{\ol{K}}
\nc{\Lbar}{\ol{L}}
\nc{\cLbar}{\ol{\cL}}
\nc{\Mbar}{\ol{M}}
\nc{\Nbar}{\ol{N}}
\nc{\Pbar}{\ol{P}}
\nc{\Qbar}{\ol{Q}}
\nc{\Rbar}{\ol{R}}
\nc{\Sbar}{\ol{S}}
\nc{\Tbar}{\ol{T}}
\nc{\Ubar}{\ol{U}}
\nc{\Vbar}{\ol{V}}
\nc{\cVbar}{\ol{\cV}}
\nc{\Wbar}{\ol{W}}
\nc{\Xbar}{{\overline X}}
\nc{\Ybar}{{\overline Y}}
\nc{\Zbar}{{\overline Z}}
\nc{\cZbar}{{\overline \cZ}}

\nc{\epsbar}{\ol{\epsilon}}
\nc{\lambar}{\ol{\lambda}}
\nc{\kapbar}{\ol{\kappa}}
\nc{\zetabar}{\ol{\zeta}}
\nc{\Zetabar}{\ol{\Zeta}}
\nc{\taubar}{\ol{\tau}}
\nc{\Taubar}{\ol{\Tau}}
\nc{\psibar}{\ol{\psi}}
\nc{\Psibar}{\ol{\Psi}}
\nc{\tpsibar}{\ol{\tpsi}}
\nc{\tPsibar}{\ol{\tPsi}}
\nc{\phibar}{\ol{\phi}}
\nc{\Phibar}{\ol{\Phi}}
\nc{\chibar}{\ol{\chi}}
\nc{\mubar}{\ol{\mu}}
\nc{\nubar}{\ol{\nu}}
\nc{\rhobar}{\ol{\rho}}
\nc{\ombar}{\ol{\om}}
\nc{\Ombar}{\ol{\Om}}
\nc{\Deltabar}{\ol{\Delta}}
\nc{\Thetabar}{\ol{\Theta}}
\nc{\xibar}{\ol{\xi}}
\nc{\Xibar}{\ol{\Xi}}

\nc{\Dthbar}{\ol{\rm D3}}

\nc{\gdot}{\dot{g}}
\nc{\pdot}{\dot{p}}
\nc{\qdot}{\dot{q}}
\nc{\rdot}{\dot{r}}
\nc{\sdot}{\dot{s}}
\nc{\tdot}{\dot{t}}
\nc{\udot}{\dot{u}}
\nc{\vdot}{\dot{v}}
\nc{\wdot}{\dot{w}}
\nc{\xdot}{\dot{x}}
\nc{\xddot}{\ddot{x}}
\nc{\ydot}{\dot{y}}
\nc{\zdot}{\dot{z}}
\nc{\yddot}{\ddot{y}}

\nc{\Udot}{\dot{U}}
\nc{\Vdot}{\dot{V}}
\nc{\Wdot}{\dot{W}}

\nc{\taudot}{\dot{\tau}}
\nc{\phidot}{\dot{\phi}}
\nc{\psidot}{\dot{\psi}}
\nc{\sinp}{s_{\phi}}
\nc{\cosp}{c_{\phi}}
\nc{\tanp}{t_{\phi}}
\nc{\spone}{s_{\phi_1}}
\nc{\cpone}{c_{\phi_1}}
\nc{\tpone}{t_{\phi_1}}
\nc{\sptwo}{s_{\phi_2}}
\nc{\cptwo}{c_{\phi_2}}
\nc{\tptwo}{t_{\phi_2}}
\nc{\spth}{s_{\phi_3}}
\nc{\cpth}{c_{\phi_3}}
\nc{\tpth}{t_{\phi_3}}
\nc{\calp}{c_{\al}}
\nc{\salp}{s_{\al}}

\nc{\csch}{{\rm csch}}
\nc{\sech}{{\rm sech}}

\nc{\cothzlami}{\coth(z-\lam_i)}
\nc{\coshzlami}{\cosh(z-\lam_i)}
\nc{\sinhzlami}{\sinh(z-\lam_i)}

\nc{\cothzlamj}{\coth(z-\lam_j)}
\nc{\coshzlamj}{\cosh(z-\lam_j)}
\nc{\sinhzlamj}{\sinh(z-\lam_j)}

\nc{\cothlamij}{\coth(\lam_i-\lam_j)}
\nc{\coshlamij}{\cosh(\lam_i-\lam_j)}
\nc{\sinhlamij}{\sinh(\lam_i-\lam_j)}

\nc{\bah}{{\mathbf {\hat{A}}}}
\nc{\bX}{{\mathbf X}}
\nc{\ba}{{\bf a}}
\nc{\bb}{{\bf b}}
\nc{\bc}{{\bf c}}
\nc{\bd}{{\bf d}}
\nc{\bg}{{\bf g}}
\nc{\bk}{{\bf k}}
\nc{\bl}{{\bf l}}
\nc{\bm}{{\bf m}}
\nc{\bn}{{\bf n}}
\nc{\bo}{{\bf o}}
\nc{\bp}{{\bf p}}
\nc{\bq}{{\bf q}}
\nc{\br}{{\bf r}}
\nc{\bs}{{\bf s}}
\nc{\bt}{{\bf t}}
\nc{\bu}{{\bf u}}
\nc{\bv}{{\bf v}}
\nc{\bw}{{\bf w}}
\nc{\bx}{{\bf x}}
\nc{\by}{{\bf y}}
\nc{\bz}{{\bf z}}
\nc{\bom}{{\bf \om}}
\nc{\bombar}{{\mathbf \ombar}}
\nc{\bPhi}{{\bf \Phi}}

\nc{\rma}{{\rm a}}
\nc{\rmb}{{\rm b}}
\nc{\rmc}{{\rm c}}
\nc{\rmd}{{\rm d}}
\nc{\rmg}{{\rm g}}
\nc{\rk}{{\rm k}}
\nc{\rml}{{\rm l}}
\nc{\rmm}{{\rm m}}
\nc{\rmn}{{\rm n}}
\nc{\rmo}{{\rm o}}
\nc{\rmp}{{\rm p}}
\nc{\rmq}{{\rm q}}
\nc{\rmr}{{\rm r}}
\nc{\rms}{{\rm s}}
\nc{\rmt}{{\rm t}}
\nc{\rmu}{{\rm u}}
\nc{\rmv}{{\rm v}}
\nc{\rmw}{{\rm w}}
\nc{\rmx}{{\rm x}}
\nc{\rmy}{{\rm y}}
\nc{\rmz}{{\rm z}}

\nc{\dal}{\dot{\al}}
\nc{\thadot}{\dot{\tha}}
\nc{\thab}{\bar{\theta}}
\nc{\thal}{\theta^{\al}}
\nc{\thdal}{\bar{\theta}^{\dal}}

\nc{\thsigthm}{\tha \sigma^m \thab}
\nc{\thsigthn}{\tha \sigma^n \thab}

\nc{\Dal}{D_{\al}}
\nc{\Ddal}{\bar{D}_{\dal}}
\nc{\CDal}{{\cal D}_{\al}}
\nc{\CDdal}{\bar{\cal D}_{\dal}}

\nc{\eq}[1]{(\ref{#1})}
\nc{\non}{\nonumber}
\nc{\Fzero}{F_{(0)}}
\nc{\Ftwo}{F_{(2)}}
\nc{\Ffour}{F_{(4)}}
\nc{\Fone}{F_{(1)}}
\nc{\Fthree}{F_{(3)}}
\nc{\Ffive}{F_{(5)}}
\nc{\Fn}{F_{(n)}}
\nc{\Fp}{F_{(p)}}

\nc{\tFzero}{\tF_{(0)}}
\nc{\tFtwo}{\tF_{(2)}}
\nc{\tFfour}{\tF_{(4)}}
\nc{\tFone}{\tF_{(1)}}
\nc{\tFthree}{\tF_{(3)}}
\nc{\tFfive}{\tF_{(5)}}
\nc{\tFn}{\tF_{(n)}}
\nc{\tFp}{\tF_{(p)}}

\nc{\Czero}{C_{(0)}}
\nc{\Ctwo}{C_{(2)}}
\nc{\Cfour}{C_{(4)}}
\nc{\Cone}{C_{(1)}}
\nc{\Cthree}{C_{(3)}}
\nc{\Cfive}{C_{(5)}}
\nc{\Cn}{C_{(n)}}

\nc{\equ}{{\rm eq}}

\nc{\vol}{{\rm vol}}
\nc{\Ainf}{A_{\infty}}
\nc{\End}{{\rm End}}
\nc{\Ext}{{\rm Ext}}
\nc{\IIB}{{\rm IIB}}
\nc{\Ad}{{\rm Ad}}
\nc{\IIA}{{\rm IIA}}
\nc{\AdS}{{\rm AdS}}
\nc{\CFT}{{\rm CFT}}
\nc{\diag}{{\rm diag}}
\nc{\Log}{{\rm Log}}
\nc{\Dslash}{\ensuremath \raisebox{0.025cm}{\slash}\hspace{-0.32cm} D}
\nc{\cDslash}{\ensuremath \raisebox{0.025cm}{\slash}\hspace{-0.32cm} \cD}
\nc{\omslash}{\om\!\!\!/}
\nc{\no}{\!:\!\!}
\nc{\ointdz}{\oint\frac{dz}{2\pi i}}
\nc{\ointdzone}{\oint\frac{dz_1}{2\pi i}}
\nc{\ointdztwo}{\oint\frac{dz_2}{2\pi i}}
\nc{\ointdzb}{\oint\frac{d\zbar}{2\pi i}}
\nc{\ointdzbone}{\oint\frac{d\zbar_1}{2\pi i}}
\nc{\ointdzbtwo}{\oint\frac{d\zbar_2}{2\pi i}}
\nc{\dz}{\frac{dz}{2\pi i}}
\nc{\dzb}{\frac{d\zbar}{2\pi i}}
\nc{\bpm}{\begin{pmatrix}}
\nc{\epm}{\end{pmatrix}}
 \nc{\bitem}{\begin{itemize}}
 \nc{\eitem}{\end{itemize}}
 \nc{\exercise}{\vskip 2mm \noindent {\bf Exercise:}}
 \nc{\definition}{\vskip 2mm \noindent {\bf Definition:}}
 
\newcommand\dd{\mathrm{d}}

\newcommand{\nn}{\nonumber \\ {} }
\newcommand\ex{\mathrm{e}}

\newcommand\qqq{\qquad\qquad}

\begin{document} 

\title{\boldmath Punctures from Probe M5-Branes and $\cN\!=\!1$ Superconformal Field Theories}

\author[a,b]{Ibrahima Bah}
\author[b,c,d]{Maxime Gabella}
\author[e]{Nick Halmagyi}

\affiliation[a]{Department of Physics and Astronomy, University of Southern California, Los Angeles, CA 90089, USA}
\affiliation[b]{Institut de Physique Th\'eorique, CEA/Saclay, 91191 Gif-sur-Yvette, France}
\affiliation[c]{Department of Mathematics, University of Hamburg, Bundesstr.~55, 20146 Hamburg, Germany}
\affiliation[d]{DESY, Theory Group, Notkestrasse 85, Bldg 2a, 22607 Hamburg, Germany}
\affiliation[e]{Laboratoire de Physique Th\'eorique et Hautes Energies,
Universit\'e Pierre et Marie Curie, CNRS UMR 7589, 
F-75252 Paris Cedex 05, France}

\emailAdd{bah@usc.edu}
\emailAdd{maxime.gabella@uni-hamburg.de}
\emailAdd{halmagyi@lpthe.jussieu.fr}

\abstract{We study probe M5-branes in $\cN\!\!=\!1$ AdS$_5$ solutions of M-theory that arise from M5-branes wrapped on a Riemann surface. Using the BPS condition from $\kappa$-symmetry, we classify supersymmetric probe M5-branes that extend along all of AdS$_5$ and intersect the Riemann surface at points. These can be viewed as punctures in the dual $\cN\!\!=\!1$ superconformal field theories. We find M5-branes that correspond to the two types of simple punctures previously studied in field theory. In addition, when the central charge is rational, we find a new class of M5-branes with a moduli space that includes two internal dimensions in addition to the Riemann surface. These new M5-branes have the essential characteristic of fractional branes, in that a single one at a generic point of its moduli space becomes multiple M5-branes at special points.}

\maketitle
\flushbottom

{\quote{\small\emph{Comme il est profond ce myst\`ere de l'Invisible! Nous ne le pouvons sonder avec nos sens mis\'erables [\ldots]
Ah! si nous avions d'autres organes qui accompliraient en notre faveur d'autres miracles, que de choses nous pourrions d\'ecouvrir encore autour de nous!}}~\footnote
{
``How profound this mystery of the Invisible is! We cannot probe it with our miserable senses [\ldots] Oh! if only we had other organs which could work other miracles in our favor, how many things we could still discover around us!''
}
\begin{flushright}\small Guy de Maupassant, ``Le Horla,'' 1887.\end{flushright}}

\section{Introduction}

M5-branes wrapped on curved manifolds yield infinite families of supersymmetric AdS solutions in various dimensions~\cite{Maldacena:2000mw}. The purpose of this article is to explore the space of AdS$_5$ solutions of M-theory which are dual to four-dimensional superconformal field theories (SCFTs) with $\N=1$ supersymmetry. Our strategy is to classify the ways in which one can deform known AdS$_5$ solutions by probe M5-branes which preserve all the superconformal symmetries. This corresponds to classifying so-called \emph{punctures} in the dual $\cN=1$ SCFTs.

The AdS$_5$ solutions that we consider describe the near-horizon region of a stack of $N$ M5-branes wrapped on a Riemann surface $\Sigma_g$.
While our calculations below are performed in the AdS backgrounds, it is enlightening to consider a UV picture, before the worldvolume theory on the M5-branes has flowed to the conformal point.  The M5-branes have the worldvolume $\RR^{1,3}\times \Sig_g$ embedded in the eleven-dimensional geometry $\RR^{1,4}\times X_3 $, where the local Calabi-Yau threefold $X_3$ is a decomposable bundle over $\Sigma_g$.
More precisely, $X_3$ consists of two line bundles $\cL_p$ and $\cL_q$ of degrees $p$ and $q$ over $\Sigma_g$:
\bea\label{LpLq}
\CC^2 \; \to\;  & \cL_p  \oplus \cL_q& \nn
&\downarrow& \\  
&\Sig_g& \nonumber
\eea
and the Calabi-Yau condition $c_1(X_3)=0$ relates the degrees to the genus as
\be
p+q=2g-2~.  
\ee
The solutions found by Maldacena and N\'u\~nez (MN)~\cite{Maldacena:2000mw} correspond to $p=0$ (or $q=0$) and  $(p,q)=(g-1,g-1)$. The former case with one trivial line bundle preserves sixteen supercharges, and the low-energy theory on the M5-branes is then an $\cN=2$ SCFT in four dimensions; the latter case preserves eight supercharges, and the low-energy theory is an $\cN=1$ SCFT. More generally, there are infinite families of such bundles $X_3$ labelled by $p$ and $q$. The corresponding supergravity solutions preserve eight supercharges and were found in~\cite{Bah:2012dg} by Bah, Beem, Bobev, and Wecht (BBBW).

The $\cN=2$ solutions admit a vast generalization found by Lin, Lunin, and Maldacena (LLM)~\cite{Lin:2004nb}, which classifies AdS$_5$ solutions of M-theory with sixteen supercharges. 
In this context, the punctures of~\cite{Witten:1997sc,Gaiotto:2009we} were interpreted in~\cite{Gaiotto:2009gz} as extra M5-branes intersecting the Riemann surface at points.
The addition of a single M5-brane, which corresponds to a \emph{simple} puncture, can be analyzed in the probe approximation.
In fact the requirement that the probe M5-brane preserves the symmetries of the $\cN=2$ solution
is enough to determine its embedding in the eleven-dimensional geometry, whose internal space is an $S^4$ fibered over $\Sigma_g$. It must extend along all of AdS$_5$ in order to preserve the $SO(2,4)$ isometries, which correspond to the four-dimensional conformal group of the dual SCFT. 
It must also preserve the $SU(2)\times U(1)$ R-symmetry, which can be made manifest by describing the $S^4$ in terms of an $S^2$ and an $S^1$ whose sizes shrink to zero at one or the other endpoint of an interval.
This is achieved when the M5-brane wraps the $S^1$ and sits at the point where the $S^2$ shrinks.
Up to a freely choosable point on $\Sig_g$, the embedding of the probe M5-brane is thus fixed:
\bea
\text{\bf{$\cN=2$ M5-brane:}} \qquad  \text{AdS$_5$} \times S^1 \times  \{\text{point $\in\Sigma_g$}\} \times  \{\text{point where $S^2$ shrinks} \}.   \nonumber
\eea

Motivated by this creation of punctures from probe M5-branes, in this paper we study supersymmetric probe M5-branes embedded in the $\cN=1$ BBBW solutions (reviewed in section~\ref{sec:AdS5}). The central calculation is performed in section~\ref{secM5}, where we analyze the BPS condition ($\kappa$-symmetry projection) for a probe M5-brane that preserves the $\cN=1$ superconformal symmetry of the dual field theories.
We find that the M5-brane always intersects the Riemann surface at a point.
It also wraps the circle $S^1_R$ dual to the $U(1)$ R-symmetry, in the middle of an interval $y$ at whose endpoints $S^1_R$ shrinks.
However, in contrast to the $\cN=2$ case, there is some freedom for the position of the M5-brane in the remaining two directions, which gives a moduli space $\cM_2$.
The embedding of a supersymmetric probe M5-brane is then:
\bea
\text{\bf $\cN=1$ M5-brane}: \qquad  \text{AdS$_5$} \times S^1_R  \times  \{\text{point $\in\Sigma_g$}\}\times  \{y=0\} \times \{\text{point $\in \cM_2$}\}  . \nonumber
\eea
There are two special points on $\cM_2$, which we call the poles.
When the M5-brane is at either pole of $\cM_2$, it preserves some additional global symmetry.
We observe that it is only for BBBW solutions with rational central charges that the M5-brane can sit at a generic point of $\cM_2$.
This leads to an interesting phenomenon, in which a single M5-brane at a generic point breaks into multiple M5-branes at the poles.
This is similar to the behavior of ``fractional branes'' in orbifold singularities~\cite{Gimon:1996rq,Douglas:1996sw}.

The presence of an extra M5-brane allows for some new configurations of BPS M2-branes ending on it along the time direction and $S^1_R$. In the internal geometry, these M2-branes have the topology of a disc since they extend away from the M5-brane towards a point where $S^1_R$ shrinks.
There are two such points, and so there is a pair of M2-branes for each M5-brane.

In section~\ref{secMN}, we compare our results to the MN solutions.
For the $\cN=1$ solutions ($p=q$), the M5-brane preserves an enhanced $U(1)_R\times U(1)_F$ everywhere on $\cM_2$, so there is no special point and the moduli space is a sphere.
In the dual SCFTs, this $S^2$ moduli space has been pointed out in~\cite{Benini:2009mz}.
For the $\cN=2$ solutions ($q=0$), the M5-brane preserves $SU(2)_R\times U(1)_R$ at one pole and $U(1)_R\times U(1)_F$ at the other. However, at any other point on $\cM_2$, only the $U(1)$ R-symmetry is preserved. This novelty prevails in all rational BBBW solutions.

While the BBBW solutions are valid for all $p$ and $q$, the corresponding quiver field theories have only been constructed for the case $p,q\ge0$ (reviewed in section~\ref{secCFT}).
There are two types of simple punctures that can be added to these quivers, and each type introduces pairs of BPS operators~\cite{Beem:2012yn,Xie:2013gma,Bah:2013aha}.
We claim that this corresponds to the two probe M5-branes at the poles of $\cM_2$,
and provide a check by showing that the volumes of the pairs of M2-branes ending on them exactly match the conformal dimensions of the pairs of BPS operators. 

On the other hand, the generic probe M5-branes, which only preserve the $U(1)$ R-symmetry, are harder to identify in field theory.

\section{AdS$_5$ solutions of M-theory}\label{sec:AdS5}

We start by reviewing the BBBW $\cN=1$ AdS$_5$ solutions that arise from $N$ M5-branes wrapping a Riemann surface~\cite{Bah:2011vv, Bah:2012dg}. 
These solutions give an interesting class of explicit examples within the general characterization of supersymmetric AdS$_5$ solutions obtained in~\cite{Gauntlett:2004zh}.
We also determine the condition under which a BBBW solution leads to a rational central charge.

\subsection{General structure}\label{probe5}

The most general solutions of the form AdS$_5 \times M_6$ with $\cN=1$ supersymmetry were analyzed in~\cite{Gauntlett:2004zh} (we adopt their conventions).
The 11d metric is of the form
\bea\label{11dmetricAdS5}
\dd s_{11}^2 &=& \ex^{2\lambda} \left[\dd s^2 (\text{AdS}_5) + \dd s^2 (M_6) \right]~.
\eea
Accordingly, the 11d gamma matrices decompose as
\bea
\Gamma^a &=& \rho^a \otimes \gamma_7~,\qqq  \text{with} \quad a=0,1,2,3,4 ~,\nn
\Gamma^{m+4} &=& 1 \otimes \gamma^m~, \qqq \text{with} \quad  m=1,2,3,4,5,6~,
\eea 
and the 11d Majorana spinor as 
\bea\label{11dMajo}
\epsilon &=& \psi_\text{AdS$_5$} \otimes \ex^{\lambda/2} \xi~.
\eea
Locally, the 6d metric splits into a 4d space $M_4$ and two directions specified by some one-forms $K^1$ and $K^2$:
\bea\label{metricM6}
\dd s^2 (M_6) &=& \dd s^2 (M_4)+ (K^1)^2 + (K^2)^2  ~. 
\eea
More explicitly, the one-forms are given by 
\bea\label{K1K2}
K^1 = \frac{\ex^{-3\lambda}} {\cos \zeta} \dd y~, \qqq 
K^2 = \frac{\cos \zeta}{3}(\dd \psi + \rho) ~,
\eea
where the function 
\bea\label{sinzetay}
\sin\zeta = 2 \ex^{-3\lambda} y 
\eea
and the one-form $\rho$ on $M_4$ are independent of~$\psi$.
An important result of~\cite{Gauntlett:2004zh} is that the spinor bilinear
\bea\label{tK2}
\tilde K^2 = \frac12 \bar \xi \gamma_\mu \gamma_7 \xi  =  \cos\zeta K^2
\eea
satisfies $\nabla_{(m} \tilde K^2_{n)}=0$, and thus defines a Killing vector field $\del/\del \psi$. 
For the solutions on which we will focus, the coordinate $\psi$ parametrizes a circle, which corresponds to the $U(1)$ R-symmetry of the dual SCFTs.

\subsection{Solutions from M5-branes on Riemann surfaces}\label{sec:BBBW}

A general class of $\cN=1$ SCFTs arising from M5-branes on a Riemann surface together with the dual AdS$_5$ solutions were found by Bah, Beem, Bobev, and Wecht (BBBW) in~\cite{Bah:2011vv, Bah:2012dg}. Here we briefly review these geometries and postpone the review of the dual SCFTs to section~\ref{secCFT}.

These solutions are labelled by the curvature $\kappa \in \{-1,0,1\}$ of $\Sigma_g$ (hyperbolic surface, torus, sphere), its genus~$g$, and the ``twist parameter'' $z$.
For $\kappa = \pm 1$,\footnote{
In this paper, we leave aside the case of the torus ($\kappa=0$), which always leads to irrational central charges.
} we can write
\bea\label{zpq}
z= \frac{p-q}{p+q}  ~, \qqq  p,q\in \ZZ~,
\eea
where, as we mention in the introduction, $p$ and $q$ are the degrees of the pair of line bundles $\cL_p\oplus \cL_q$ over $\Sigma_g$ and satisfy the Calabi-Yau condition
\bea\label{pqCYcondition}
p+q = 2g-2~.
\eea
For the sphere, the range of $z$ should be restricted to $|z|>1$.

The metric of the $\N=1$ BBBW solutions~\cite{Bah:2012dg} is of the form AdS$_5\times M_6$ as in~\eqref{11dmetricAdS5}, and $M_6$ consists of a 4-manifold $M_4$ together with the two directions $K^1$ and $K^2$ as in~\eq{metricM6}. The metric on $M_4$ splits into a Riemann surface $\Sigma_g$ with coordinates $\{x_1,x_2\}$, an interval with coordinate~$w$,\footnote
{
This coordinate $w$ was called $q$ in~\cite{Bah:2012dg}.
}
and a circle with coordinate $\chi$ fibered over $\Sigma_g$:
\bea \label{M4met}
\dd s^2 (M_4) &=& \ex^{2\nu}   \ex^{2A} \left(\dd x_1^2 + \dd x_2^2\right) + \left(P^1\right)^2 + \left(P^2\right)^2~,
\eea
with the one-forms $P^1$ and $P^2$ defined as
\bea\label{P1P2}
P^1 &=& \frac{\cos\zeta}{\sqrt{k(w)}} \left(\dd w + 12 \ex^{-6\lambda}\frac{k(w)}{\cos^2\zeta} y\dd y \right)  ~, \nn 
P^2 &=& 2\kappa(1-g) \ex^{2\nu}  \cos \zeta \frac{\sqrt{k(w)}}{w} \left(\dd \chi +  V\right)  ~.
\eea 
The function $A$ satisfies the Liouville equation $ (\partial_{x_1}^2+\partial_{x_2}^2)   A=-\kappa \ex^{2A}$, and can be written as
\begin{equation}
\ex^{A} = \frac{2}{1+\kappa(x_1^2 +x_2^2)}~.  
\end{equation}
The remaining functions are given by
\bea\label{zetak}
k(w) &=& \frac{(w_+-w)(w-w_-)}{w_++w_-} ~,    \nn
 \cos^2\zeta &=& \frac{6w ( 6w_+ w_- - y^2)}{6w(6w_+ w_- -y^2) + 4(w_++w_-) y^2}~,\nn
\ex^{-6\lambda} &=& \frac{w_++w_-}{6w (6w_+w_--y^2)}\cos^2\zeta~.
\eea
We see that in order for the metric to be well-defined, the interval coordinates must take value between the zeroes of $k(w)$ and $\cos\zeta$, that is $w\in [w_-,w_+]$
and $y\in[-y_0,y_0]$ with 
\bea
y_0 = \sqrt{6w_+w_-}~.
\eea
The constants appearing in the metric can be expressed as
\bea
 w_\pm = \frac1{12} \frac{6\ex^{2\nu} \pm  \kappa z}{6\ex^{2\nu} - \kappa} ~ , \qqq  6\ex^{2\nu} = - \kappa + \sqrt{1 + 3 z^2} ~. \label{acpart}
\eea
Note that $w_\pm\ge 0$ and $w_+ +w_-= 36 w_+w_-$. We take $w_+\ge w_-$, which requires $\kappa z \ge 0$.

\begin{figure}[t]
\centering
\includegraphics[width=6.5in]{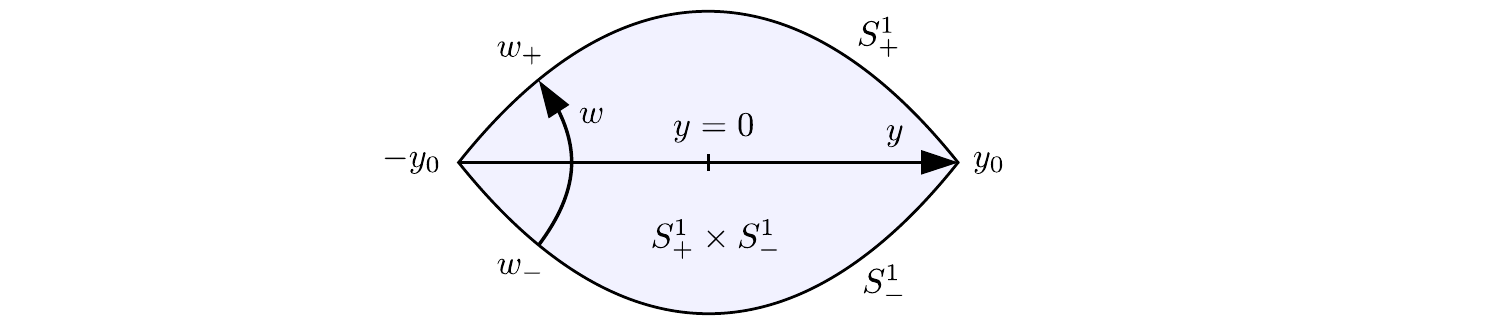}
\caption{Region delimited by $-y_0\le y\le y_0$ and $w_-\le w\le w_+$. The $w$-interval shrinks as it approaches the extremities of the $y$-interval.
At a generic point, the torus $S^1_+\times S^1_-$ parametrized by $\{\phi_+,\phi_-\}$ is finite, but on the upper (lower) boundary with $w=w_\pm$ the circle $S^1_\mp$ shrinks, leaving only one circle $S^1_\pm$. At the extremities $y=\pm y_0$, both circles shrink.}
\label{eye}
\end{figure}
\begin{figure}[t]
\centering
\includegraphics[width=6.5in]{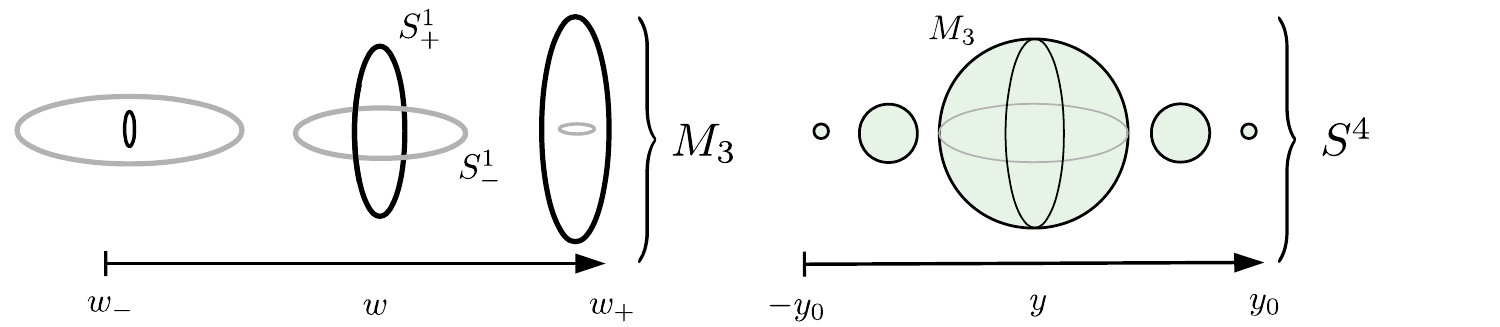}
\caption{The circles $S^1_\pm$ and the $w$-interval give a 3-manifold $M_3$, which itself combines with the $y$-interval to give an $S^4$.}
\label{S3S4}
\end{figure}

The one-forms determining the fibrations of the $\psi$ and $\chi$ directions are given by 
\bea\label{rhoV}
\rho &=&  \kappa(1-g)\ex^{2\nu} \frac{2 w - w_+-w_-}{w(w_++w_-)}  (\dd  \chi +  V) + (2-2g)  V  ~,\nn
V &=& \frac{\kappa}{1-g} \frac{x_1 \dd x_2-x_2 \dd x_1 }{1+ \kappa (x_1^2+x_2^2)}~.  
\eea
The presence of $\dd \chi$ in the connection $\rho$ implies that the coordinates $\psi$ and $  \chi$ mix in the fibration defined by $K^2$ in~\eqref{K1K2}. 
We can make the following change of variables to angle coordinates $\phi_\pm$ which fiber independently over $\Sigma_g$:\footnote
{
For reasons that will become clear below, we switched the notation $\phi_+ \leftrightarrow \phi_-$ with respect to~\cite{Bah:2012dg}.
}
\bea\label{psichi2phipm}
\psi = \phi_+ + \phi_- ~, \qqq
  \chi =  \frac{\kappa \ex^{-2\nu}}{1-g}\frac{w_++w_-}{w_+ -w_-} \left(w_- \phi_+ -w_+ \phi_-\right)~.
\eea
The part of the metric describing the fibers then diagonalizes: 
\bea\label{metricphipm}
\left(K^2\right)^2 +  \left(P^2\right)^2  = 4\frac{w_++w_-}{w(w_+-w_-)}\cos^2\zeta  & \bigg[ & w_-(w-w_-)  \left(\dd\phi_+ -  {p}  V\right)^2 \nn
&& +  w_+(w_+-w)  \left(\dd\phi_- - {q}  V\right)^2 \bigg]~.
\eea
We see that the circles $S^1_\pm$ with coordinates $\phi_\pm$ shrink at $w=w_\mp$, respectively. 
However, the metric is smooth if we take the periods of $\phi_\pm$ to be $2\pi$.
In addition, both circles shrink at $\cos\zeta=0$, that is at $y=\pm y_0$.
In contrast, the Killing vector fields
\bea\label{Killingvf}
\frac{\partial}{\partial\psi} &=& \frac{w_+}{w_++w_-} \del_+ +  \frac{w_-}{w_++w_-} \del_-  ~,  \nn
\frac{\partial}{\partial \chi}  &=& \kappa(1-g) \ex^{2\nu} \frac{w_+-w_-}{(w_++w_-)^2} \left( \del_+ - \del_- \right)
\eea
have the norms
\bea
\left\| \frac{\partial}{\partial\psi}  \right\|^2 &=& \frac{ \cos^2\zeta}{9}~, \nn
\left\| \frac{\partial}{\partial \chi}  \right\|^2 &=& 4 \left( 1-g \right)^2 \ex^{4\nu} \cos^2\zeta  \frac{(w_+- w_-)^2(w_+ + w_- -w) }{w(w_+ + w_-)^3}~,  \label{cyclenorms}
\eea
which only vanish for $\cos\zeta =0$.
Note finally from the expression for $P^1$ in~\eq{P1P2} that the $k$-interval shrinks at the two endpoints $\pm y_0$ of the $y$-interval. The $\{y,w\}$-region thus has the shape of an eye, as we sketched in figure~\ref{eye}.
In fact, topologically, the coordinates $\{\phi_+,\phi_-,w\}$ describe a compact 3-manifold $M_3$, which shrinks at the end of the $y$-interval, thus forming a 4-sphere (see figure~\ref{S3S4}).
%

\subsection{Rational central charges}\label{RationalCC}

We now pause to analyze the condition for the internal volume of a BBBW solution, or equivalently the central charge of the dual SCFT, to be rational. 
Indeed, it will be important in the next section that the superconformal R-symmetry is $U(1)$ rather than $\RR$.
This is similar in spirit to the condition for the $Y^{p,q}$ solutions of IIB supergravity to have rational volumes \cite{Gauntlett:2004yd}. 

The central charges for these SCFTs were computed exactly in \cite{Bah:2012dg} and are given in the large $N$ limit by 
\bea
a=c= (1-g) \frac{1-9  z^2+\kappa (1+3z^2)^{3/2}}{48 z^2} N^3~.
\eea  
This expression is rational when the parameter $z$ defined in~\eqref{zpq} satisfies the condition
\bea\label{rationalcc}
\sqrt{1+3z^2}   \in \QQ~.
\eea
The general solution can be written as
\bea\label{zSol}
z&=& \frac{2n_1 n_2}{n_2^2-3n_1^2}~, \qqq n_1,n_2\in \ZZ~.
\eea
In terms of the Chern numbers $p$ and $q$ that satisfy~\eq{pqCYcondition}, this rationality condition becomes
\be
(3 n_1^2 - n_2^2)(p-q) + 4n_1n_2 (g-1)=0~. \label{rational}
\ee 
So for each pair of integers $(p-q,g)$ for which there exists a solution to this condition, there is an SCFT with rational central charge. 

We will not classify all such integer pairs but one strategy to solve \eq{rational} is to first solve the Pell equation
\be\label{PellEqn}
n_2^2-3 n_1^2 =n_0
\ee
for $n_0\in \ZZ$. Then for each such solution $(n_1,n_2)$, we have a rational SCFT as long as 
\be\label{pmqrational}
p-q=\frac{4n_1n_2(g-1)}{n_0}
\ee
is an integer. This indicates that rational SCFTs are abundant. Indeed, the Pell equation \eq{PellEqn} with $n_0=1$ is known to have infinitely many integer solutions and for each such solution $p-q$ is an integer.%
\footnote{Other simple choices for $n_0$ which would ensure that $p-q$ is an integer for arbitrary $g$ are $n_0\in\{-1,\pm2,\pm4\}$. However for $n_0\in\{-1,\pm2,-4\}$ it is known that \eq{PellEqn} has no solutions and that $n_0=4$ gives the same integers $(p-q,g)$ as $n_0=1$.}
There is thus an infinite family of rational SCFTs for any $g\neq 1$. Another family of solutions can be obtained from solutions to \eq{PellEqn} with $n_0=\pm (g-1)$.

\section{Supersymmetric probe M5-branes}\label{secM5}

We now classify all the embeddings for a probe M5-brane that wraps AdS$_5$ and preserves the supersymmetries in the $\cN=1$ BBBW backgrounds.
Such embeddings saturate a BPS bound written in terms of a $\kappa$-symmetry projector.
The M5-brane is calibrated by the Killing one-form $\tilde K^2$, and wraps the $S^1_R$ corresponding to the $U(1)$ R-symmetry of the dual SCFT.
We discover that there is a two-dimensional moduli space $\cM_2$, at the poles of which the M5-brane preserves an additional $U(1)$ flavor symmetry. 
The M5-brane turns into multiple M5-branes when it reaches the poles, in a way reminiscent of fractional branes at orbifold fixed points.
We also study BPS M2-branes ending on the M5-brane and compute the conformal dimensions of the dual operators (which we will construct in section~\ref{secCFT}).

\subsection{BPS condition}

The requirement of $\kappa$-symmetry leads to a BPS bound on a supersymmetric probe M5-brane \cite{Becker:1995kb}.
A configuration that preserves some supersymmetry satisfies $\cP_-\epsilon=0$, where
$\epsilon$ is a Majorana spinor of 11d supergravity satisfying the Killing spinor equation, and $\cP_-$ is a $\kappa$-symmetry projector.
Explicitly, we have $\cP_- = (1-\tilde \Gamma)/2$ with~\footnote
{
We omit the contribution of the worldvolume flux since there won't be any in the cases of interest in this paper.
}
\bea
\tilde \Gamma &=& \frac1{5! \cL_\text{M5}}   \Gamma_0\Gamma^{N_1\cdots N_5} \varepsilon_{N_1\cdots N_5} |_{\text{M5}}~,
\eea
where $\cL_{\text{M5}} = \sqrt{g_\text{M5}}$ is the Dirac-Born-Infeld Lagrangian on the M5-brane
($g_\text{M5}$ is the determinant of the induced metric),
and the subscript $|_{\text{M5}}$ denotes the pullback to the worldvolume of the M5-brane.
We can thus write the following BPS bound:
\bea
\| \cP_- \epsilon \|^2 = \epsilon^\dag \cP_- \epsilon \ge 0~,
\eea
which is saturated if and only if the probe M5-brane is supersymmetric. 
We can rewrite this bound as
\bea\label{Bound5}
 \epsilon^\dag \epsilon \cL_{\text{M5}} \vol_5  &\geq&   \nu_5 |_\text{M5} ~,
\eea
where $\vol_5$
is the volume form on the spatial part of the worldvolume of the M5-brane,
and the five-form $\nu_5$ is defined as the bilinear
\bea
\nu_5 = \bar \epsilon \Gamma_{(5)}\epsilon~,
\eea
with $\bar \epsilon = \epsilon^\dag \Gamma_0$ and $\Gamma_{(n)} = \frac1{n!} \Gamma_{N_1\cdots N_n}\dd X^{N_1} \wedge \cdots\wedge \dd X^{N_n}$.

There is a similar BPS condition for a supersymmetric probe M2-brane:
\bea\label{Bound2}
\epsilon^\dag \epsilon \cL_\text{M2} \vol_2  \geq  \mu_2 |_\text{M2} ~, \qqq \text{with}\qquad \mu_2 = \bar\epsilon \Gamma_{(2)} \epsilon~.
\eea

\subsection{Probe M5-Branes}\label{AdS5probeM5}

We now find the possible embeddings for a probe M5-brane in the BBBW geometries that preserve the superconformal symmetries of the dual SCFTs.
First, the M5-brane has to extend along all of AdS$_5$ in order to preserve its $SO(2,4)$ isometry group, dual to the four-dimensional conformal group. 
This also means that there cannot be any three-form flux $H$ on the worldvolume of the M5-brane, since it would have to extend along some directions in AdS$_5$, thus breaking $SO(2,4)$. 
What remains to be determined is the internal one-cycle that the M5-brane can wrap without breaking the $U(1)$ R-symmetry.

Taking $\{\sigma^1,\sigma^2,\sigma^3,\sigma^4\}$ to be spatial coordinates in AdS$_5$, and $\tau=\sigma^5$ to be the proper-length parameter along the one-cycle in $M_6$,
we find that the five-form on the RHS of the BPS condition~\eqref{Bound5} becomes
\bea
\nu_5|_\text{M5} &=&  2 e^{7\lambda}  \sqrt{g_\text{AdS$_5$}}\|\psi_\text{AdS$_5$}\|^2 \dd \sigma^{1234}   \wedge  \tilde K^2|_{\dd\tau} ~.
\eea
We see that the probe M5-brane is calibrated by the Killing one-form $\tilde K^2$ defined in~\eqref{tK2}.
On the other hand, we have $\epsilon^\dag \epsilon= 2 \ex^{\lambda} \|\psi_\text{AdS$_5$}\|^2$ from~\eq{11dMajo}, and the DBI Lagrangian splits as $\cL_\text{M5} = \ex^{6\lambda} \sqrt{g_\text{AdS$_5$}} \sqrt{g_{\tau \tau}}$. 
The BPS bound is then saturated when
\bea\label{BPStK2}
\sqrt{g_{\tau\tau}} &=& \widetilde{K}^2 |_{\dd\tau}~.
\eea
The component $g_{\tau \tau}$ of the induced metric can be obtained from the metric on $M_6$ given in~\eqref{metricM6} and takes the schematic form
\bea\label{gtautau}
g_{\tau \tau} &=& \left( \dd s_4|_{\dd\tau}\right)^2 + \left( K^1|_{\dd\tau}\right)^2  + \left( K^2|_{\dd\tau}\right)^2 ~.
\eea 
Recalling that $\widetilde{K}^2 = \cos\zeta K^2$, we can rewrite the BPS condition~\eqref{BPStK2} as
\bea
\left( \dd s_4|_{\dd\tau}\right)^2 + \left( K^1|_{\dd\tau}\right)^2  +\sin^2 \zeta\left( K^2|_{\dd\tau}\right)^2 =0~.
\eea
Each term must vanish separately, which gives the following BPS conditions:
\bea\label{Ads5BPSM5}
\dd s_4|_{\dd\tau} =0~, \qqq 
K^1|_{\dd\tau}=0~, \qqq
 \sin  \zeta K^2|_{\dd\tau} =0~.
\eea
The only way to satisfy these conditions without making the induced metric~\eqref{gtautau} vanish is to set
\bea\label{sinzeta0}
\sin \zeta =0~.
\eea
Recalling the definition~\eq{sinzetay}, we conclude that this requires that the M5-brane sits right in the middle of the $y$-interval:
\bea\label{BPSy0}
y=0~.
\eea
The second condition in~\eq{Ads5BPSM5} is then automatically satisfied, since $K^1$ is proportional to $\dd y$. 

Given the form of the BBBW metric~\eq{M4met}, the remaining BPS condition $\dd s_4|_{\dd\tau}=0$ can be expressed more explicitly as
\bea
\dd x_i|_{\dd\tau}=0~, \qqq P^1|_{\dd\tau}=0~, \qqq P^2|_{\dd\tau}=0~.
\eea
The first condition implies that $\{x_1,x_2\}$ are independent of $\tau$, and so the M5-brane intersects the Riemann surface at a constant point.
Adding an M5-brane can thus be thought of as creating a simple puncture on the Riemann surface.
In turn, from the expression for $P^1$ in~\eq{P1P2}, we see that the second condition implies that the M5-brane must sit at a constant point $w_0$ on the $w$-interval.

We now analyze several ways to satisfy the last condition for $P^2$, which differ by the position of the M5-brane on the $w$-interval.
If the M5-brane sits at one of the endpoints $w_+$ or $w_-$ we call it {\bf M5$_+$} or {\bf M5$_-$}, while if it sits at a generic point $w_0$ we call it {\bf M5$_0$}. 

\paragraph{M5$_\pm$:}
We see from~\eq{P1P2} that $P^2$ vanishes where $k(w)=0$, that is at the endpoints $w_\pm$ of the $w$-interval.
Recall that at $w_\pm$ the circle $S^1_{\mp}$ shrinks. So the M5-brane wraps the circle $S^1_{\pm}$, which corresponds to the $U(1)$ R-symmetry of the dual SCFT.
The isometry of the shrunken circle $S^1_{\mp}$ is also preserved by these configurations, and corresponds to a $U(1)$ flavor symmetry.
We summarize the two embeddings for M5-branes at $w_\pm$ as
\bea
\mathbf{\text{\bf M5}_\pm :} \qquad  \text{AdS$_5$} \times S^1_\pm \times  \{y=0\}\times \{w_\pm, S^1_\mp=0 \}  \times  \{\text{point $\in\Sigma_g$}\}. \nonumber
\eea

\paragraph{M5$_0$:}
At a generic point $w_0$ different from the endpoints $w_\pm$, the BPS condition $P^2|_{\dd\tau}=0$ imposes that $\chi$ is independent of $\tau$: $\chi = \chi_0$ with $\chi_0$ constant.  Via the change of variables~\eq{psichi2phipm} this gives
\be\label{bpsM50}
w_-\frac{\dd\phi_+}{\dd\tau} = w_+\frac{\dd\phi_-}{ \dd\tau} ~.
\ee 
This describes a curve on the torus $S^1_+\times S^1_-$, or equivalently a straight line in the $\{\phi_\pm\}$-plane with slope $w_+/w_-$.
If the slope is irrational, the curve covers the torus densely and never closes, and is thus not an appropriate cycle for the M5-brane to wrap.
We impose that the slope is rational, which gives the condition 
\begin{equation} \label{wpmmpm}
\frac{w_+}{w_-} = \frac{2z+\kappa \sqrt{1+3 z^2}}{z-1}  = \frac{m_+}{m_-} \quad \in \QQ~,  
\end{equation} 
where we have introduced a pair of positive coprime integers $m_+$ and $m_-$.
This condition is equivalent to the condition for rational central charges
analyzed in section~\ref{RationalCC}, and $m_\pm$ can be expressed in terms of the integers $n_1$ and $n_2$ introduced in~\eq{zSol}:\footnote
{
These expressions should be reduced when appropriate to obtain coprime $m_\pm$.
}
\bea \label{defmpm}
m_\pm &=& | (1-2\kappa) n_1 \pm  n_2 |  ~,\qqq\;\;\, \text{when}\quad 3 n_1^2-n_2^2 >0~,  \nn
m_\pm &=& |(1+2\kappa) n_1  \pm n_2 |   ~,\qqq\;\;\, \text{when}\quad 3 n_1^2-n_2^2 < 0~.
\eea
Note that for $\kappa=-1$, we have the bound $1 \le w_+/w_- \le 2+\sqrt3$,
which implies that the angle of inclination of the curve in the $\{\phi_\pm\}$-plane obeys the bound $\pi/4\le \gamma \le 5\pi/12$;
for $\kappa=1$ we have instead $w_+/w_- \ge 2+\sqrt3$.  
We parametrize the curve as
\begin{equation}
\phi_+(\tau) = \phi_+^0 + m_+ \tau~, \qqq \phi_-(\tau) = \phi_-^0 + m_- \tau~,
\end{equation}  
where $\phi_\pm^0$ are constants indicating the initial point of the curve.
This closed curve with slope $m_+/m_-$ wrapped by the M5-brane corresponds to the $U(1)$ R-symmetry ($\del_\tau$ is proportional to the Killing vector field $\del_\psi$~\eq{Killingvf}), so we denote it by $S^1_R\left(m_+/m_-\right)$.
On the other hand, the M5-brane breaks the second $U(1)$ symmetry. 
A probe M5-brane {\bf M5$_0$} thus has the following two-parameter family of supersymmetric embeddings:
\bea
\text{\bf M5$_0(w_0,\chi_0)$}: \qquad  \text{AdS$_5$} \times S^1_R\big(\tfrac{m_+}{m_-}\big) \times  \{y=0\} \times \{w_0, \chi_0 \}  \times  \{\text{point $\in\Sigma_g$}\} . \nonumber
\eea
We will study the moduli space parametrized by $\{w_0,\chi_0\}$ in more detail below.

\

In summary, we have  found two types of supersymmetric probe M5-branes in the BBBW geometries: {\bf M5$_\pm$} sit at the endpoints $w_\pm$ and preserve $U(1)_R\times U(1)_F$, while {\bf M5$_0$} can sit at any point $w_0$ on the $w$-interval and only preserves $U(1)_R$.

Let's consider what happens when we take an {\bf M5$_0$} wrapping $S^1_R\left(m_+/m_-\right)$ at a generic point $w_0$ and move it towards one of the endpoints $w_\pm$.
Since the circle $S^1_\mp$ shrinks at $w_\pm$, we end up with an M5-brane wrapping $m_\pm$ times the circle $S^1_\pm$, or equivalently with a number $m_\pm$ of coincident M5-branes {\bf M5$_\pm$}.
Reversely, we need a number $m_\pm$ of M5-branes {\bf M5$_\pm$} at $w_\pm$ in order to move them off along the $w$-interval, which turns them into a single {\bf M5$_0$}.
This is the characteristic feature of fractional branes on orbifolds~\cite{Douglas:1996sw,Gimon:1996rq}. 

\paragraph{Moduli space:}
We can understand this phenomenon better by studying the moduli space of a probe M5-brane. 
Since the M5-brane is calibrated by $K^2$ and sits at the point $y=0$, its moduli space is a four-dimensional space%
\footnote{
Only when the Killing vector field $\del_\psi$, dual to $K^2$, has a closed orbit does it make sense to talk about the base space $\cM_4$, and this is equivalent to the central charge of the dual SCFT being rational. This is an identical phenomenon to the well-known analysis on Sasaki-Einstein manifolds, see~\cite{Gauntlett:2004yd} for a discussion.
} 
$\cM_4$, whose metric can be found by evaluating $\dd s^2(M_4)$ in~\eq{M4met} at $y=0$. The space $\cM_4$ consists of the Riemann surface $\Sig_g$ and a two-dimensional space $\cM_2$ fibered over it:
\bea\label{M4M2}
\cM_2 \; \to\;  & \cM_4& \nn
&\downarrow& \\  
&\Sig_g&. \nonumber
\eea
While the full six-dimensional space $M_6$ in~\eq{metricM6} is smooth, for generic parameters $(p,q)$ the fiber $\cM_2$ itself is not smooth. 
We obtain the metric on $\cM_2$ by setting $y=0$ and $V=0$ in $ (P^1)^2+(P^2)^2$:
\bea
\dd s^2 (\cM_2) = \frac{\dd w^2}{k} + 4  \frac{k}{w^2} \left(\frac{w_++w_-}{w_+-w_-}\right)^2\dd \hchi^2 ~,
\eea
where we defined 
\be\label{chiphi}
\hchi= w_- \phi_+ - w_+ \phi_-~.
\ee 
Much like a two-sphere, $\cM_2$ consists of an interval $w$ and an angle $\hchi$ that shrinks at the endpoints $w_\pm$, but it can have conical singularities there.\footnote
{
The three-dimensional space given by $\{ K^2,P^1,P^2\}$ (at fixed $y$) appears to be a Seifert fiber space.
} 
\begin{figure}[t]
\centering
\includegraphics[width=6.5in]{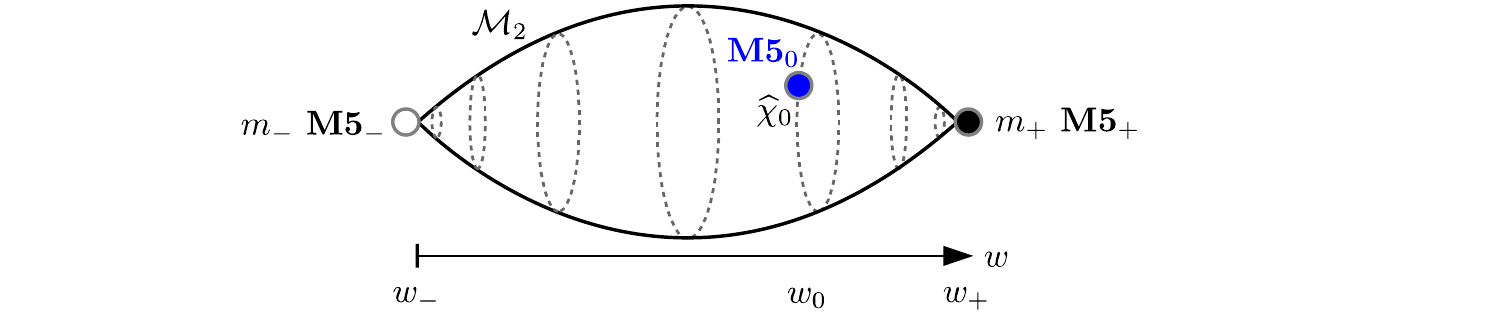}
\caption{Moduli space $\cM_2$ for a supersymmetric probe M5-brane, with coordinates $\{w_0,\hchi_0\}$. When a single {\bf M5$_0$} wrapping $S^1_R\left(m_+/m_-\right)$ at $w_0$ moves to the endpoint $w_\pm$ it becomes a number $m_\pm$ of M5-branes of type {\bf M5$_\pm$}.}
\label{S2moduli}
\end{figure}
Expanding the metric at $w_\pm$, we find
\bea
w=w_+: &&\qqq \dd s^2 (\cM_2) \sim \frac{\dd w^2}{w_+-w} + \frac{4(w_+ - w)}{w_+^2}\dd\hchi^2  \label{M2wplus}~, \nn
w=w_-: &&\qqq \dd s^2 (\cM_2) \sim \frac{\dd w^2}{w-w_-} +  \frac{4 (w - w_-)}{w_-^2} \dd\hchi^2 \label{M2wminus}~.
\eea
Requiring that the ranks of the conical singularities are integers determines the period of $\hchi$ to be
\bea
0 \leq \hchi \leq 2\pi \frac{w_+}{m_+}  = 2\pi \frac{w_-}{m_-} = 2\pi \frac{m_++m_-}{36 m_+ m_-} ~,
\eea
where we used that $w_+ +w_-= 36 w_+w_-$.
By matching the volumes of $M_6$ computed with $\phi_\pm$ and with $\{\psi,\hchi \}$, we then assign the following periodicity to $\psi$:
\be
 0 \leq \psi \leq 2\pi (m_+ + m_-)  ~.
\ee
In conclusion, we find conical singularities of the form $\RR^2 /\mathbb{Z}_{m_\pm}$ at the endpoints $w_\pm$.
This is consistent with the fractional behavior of the probe M5-brane {\bf M5$_0$}, which turns into a number $m_\pm$ of M5-branes {\bf M5$_\pm$} at $w_\pm$.

\subsection{M2-branes ending on M5-branes}\label{sec:M2onM5}

The presence of an extra M5-brane allows for new configurations of probe M2-branes ending on it.
We consider a probe M2-brane which moves along a geodesic in AdS$_5$ and wraps an internal surface whose boundary is the circle wrapped by the M5-brane. This corresponds to chiral BPS operators in the dual SCFT, as we will illustrate in section~\ref{secCFT}.

The calibration two-form appearing in the BPS condition~\eqref{Bound2} is given by (see~\cite{Gauntlett:2006ai})
\bea
\mu_2 &=& 2 \ex^\lambda   \psi_\text{AdS$_5$}^\dag \rho_0 \psi_\text{AdS$_5$} Y'~,
\eea
where 
\bea
Y'  = \frac12 \xi^\dag \gamma_7 \gamma_{(2)} \xi= K^1 \wedge K^2 - \sin \zeta J~,
\eea 
and $J$ is the $(1,1)$-form of the local $SU(2)$-structure on $M_4$.
The AdS$_5$ part of the BPS condition reads
\bea
\|\psi_\text{AdS$_5$}\|^2   = \psi_\text{AdS$_5$}^\dag \rho_0 \psi_\text{AdS$_5$}  ~,
\eea
which implies that the M2-brane is at the center of AdS$_5$ (see~\cite{Gauntlett:2006ai} for more details).
The internal part of the BPS condition is
\bea\label{bpsM2internal}
\sqrt{g_\text{M2}} = \left[ K^1 \wedge K^2 - \sin \zeta J \right] \big|_\text{M2}~,
\eea
where $g_\text{M2}$ is the determinant of the metric on the spatial worldvolume of the M2-brane induced from $\dd s^2(M_6)$.
The $(1,1)$-form $J$ is given in~\cite{Bah:2012dg} as
\bea
J = \ex^{2\nu}\ex^{2A} \dd x_1 \wedge \dd x_2 +  P^1 \wedge P^2~. 
\eea
The first term means that the M2-brane can wrap the Riemann surface $\Sigma_g$, as was analyzed in~\cite{Bah:2012dg}.
Here, we are interested in an M2-brane that ends on the circle $S^1_R$ wrapped by the probe M5-brane and extends away from it towards a point where its worldvolume closes (it thus has the topology of a cup or a disc).
The embedding of the probe M2-brane is then
\bea
\text{\bf{BPS M2-brane:}} \quad  \{\text{$ t \in$ AdS$_5$}\} \times S^1_R \times \{\text{interval $\in\{y,w\}$} \}  \times  \{\text{pt $\in S^1$}\}  \times  \{\text{pt $\in\Sigma_g$}\}  ~. \nonumber
\eea

\begin{figure}[tbh]
\centering
\includegraphics[width=6.5in]{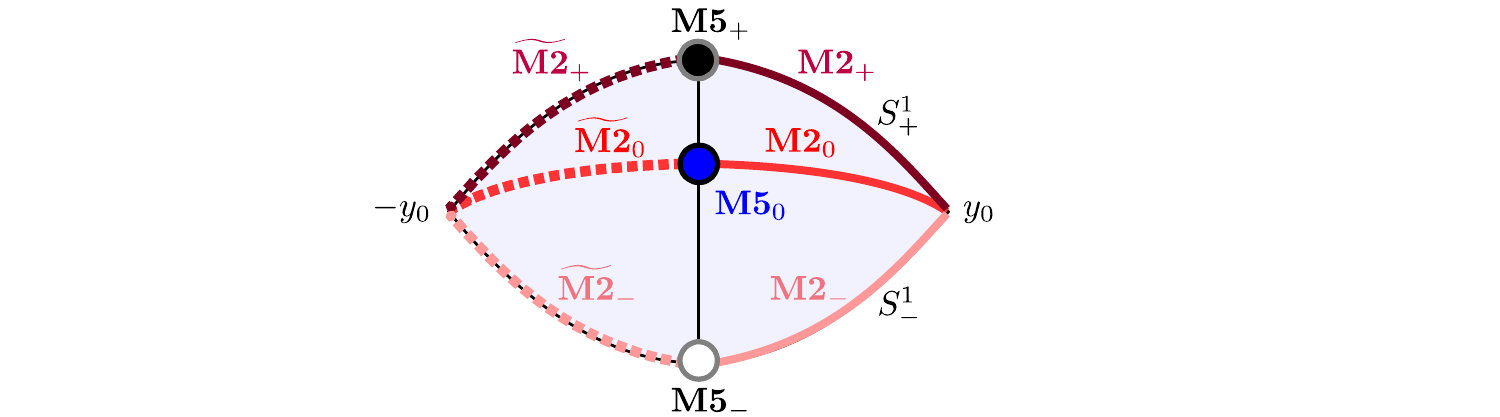}
\caption{Positions of the probe M5-branes on the vertical axis $y=0$: {\bf M5$_+$} at $w_+$ (black dot), {\bf M5$_0$} at $w_0$ (blue dot), and {\bf M5$_-$} at $w_-$ (white dot). 
Colored curves indicate pairs of M2-branes $\{\text{\bf M2}, \widetilde{ \mathrm{ \bf M2}} \}$ ending on the M5-branes and extending towards $\pm y_0$ where all circles shrink.}
\label{eyeM5M2}
\end{figure}

We parametrize the circle $S^1_R$ by $\tau$ as before, and the interval in the $\{y,w\}$-region by a second worldvolume coordinate $\sigma$: 
\bea
\phi_\pm = \phi_\pm(\tau)~, \qqq \{y,w\} = \{y,w\}(\sigma)~.
\eea
Recalling that $S^1_R$ is determined by the condition $P^2|_{\dd\tau}=0$, we find that the BPS condition~\eq{bpsM2internal} for the M2-brane simply reads
\bea\label{M2P10}
P^1|_{\dd\sigma} =0~.
\eea
This implies that an M2-brane ending on {\bf M5$_+$} or {\bf M5$_-$} extends from $y=0$ to either extremity $y=\pm y_0$, along the contour of the eye-shaped $\{y,w\}$-region (see figure~\ref{eyeM5M2}).
We denote the pair of M2-branes for each M5-brane by $\{\text{\bf M2}_\pm, \widetilde{ \mathrm{ \bf M2}}_\pm \}$.
Similarly, an M2-brane ending on {\bf M5$_0$} extends along the curve
\begin{equation}
y^2 = y_0^2 - C (w_+-w)^\frac{-w_+}{w_+-w_-} (w-w_-)^\frac{w_-}{w_+-w_-} ~,
\end{equation} 
which also has two branches ending at either $y=\pm y_0$.
Here $C\ge0$ an integration constant, which can be adjusted so that the curve meets the axis $y=0$ at the point $w=w_0$ where the M5-brane is.
We denote such a pair of M2-branes by $\{\text{\bf M2}_0, \widetilde{ \mathrm{ \bf M2}}_0 \}$.

\paragraph{Conformal dimensions:}
The volume of the submanifold $M_2\subset M_6$ wrapped by a supersymmetric probe M2-brane (dressed by suitable powers of the warp factor) corresponds to the conformal dimension of the dual BPS operator~\cite{Gauntlett:2006ai}:
\begin{equation}
\Delta(M_2) = \frac{L^3}{4\pi^2 \ell_{11}^3} \int \ex^{3\lambda} \vol_{M_6}(M_2)~, \qquad   
\end{equation} 
where $\vol_{M_6}(M_2)$ is the volume form of $M_2$ induced from $\dd s^2(M_6)$.
Note that we have reintroduced the overall length scale $L^2$ in front of the metric~\eq{11dmetricAdS5}, which was so far set to $1$.
The quantization condition for the four-form flux $G_4$ relates $L$ to the number $N$ of original M5-branes wrapping the Riemann surface $\Sigma_g$:
\bea\label{NL3}
N =  \frac{L^3}{9\pi \ell_{11}^3} (6y_0)^3~.
\eea

In the case at hand of an M2-brane wrapping a cup $M_2$ that ends on an M5-brane, this becomes
\bea
\Delta(M_2) &=&  \frac{9N}{4\pi (6y_0)^3} \int_{M_2} \ex^{3\lambda} K^1 \wedge K^2   \nn
&=& \frac{3N}{4\pi (6y_0)^3} \int_{M_2}  \dd y (\dd\psi + \rho)~.
\eea
For a pair of M2-brane $\{\text{\bf M2}_\pm, \widetilde{ \mathrm{ \bf M2}}_\pm \}$ ending on an M5-brane {\bf M5$_\pm$} at $w=w_\pm$ we find
\bea\label{dimAprobe}
\Delta[\text{\bf M2}_\pm ]  =  \Delta[ \widetilde{ \mathrm{ \bf M2}}_\pm]  = \frac{3N}{2} \frac{w_\mp}{w_++w_-} ~,
\eea
while for a pair $\{\text{\bf M2}_0, \widetilde{ \mathrm{ \bf M2}}_0 \}$ ending on an M5-brane {\bf M5$_0$} at $w=w_0$ we find
\bea \label{Deltaw0}
\Delta[\text{\bf M2}_0 ]  =   \Delta[ \widetilde{ \mathrm{ \bf M2}}_0]  =  \frac{3N}{4}  \frac{m_+ w_- + m_- w_+}{w_++w_-} ~.
\eea
We see that the conformal dimensions do not depend on the position $\{w_0,\chi_0\}$ of the M5-brane on its moduli space.

Observe however that using the relation~\eq{wpmmpm} between $w_\pm$ and $m_\pm$ we get
\bea\label{DimM2}
\Delta[\text{\bf M2}_0 ]  = m_\pm \Delta[\text{\bf M2}_\pm ] ~.
\eea
We interpret this as meaning that a single probe M2-brane ending on a single {\bf M5$_0$} at a generic position $w_0$ becomes $m_\pm$ M2-branes ending on $m_\pm$ M5-branes of type {\bf M5$_\pm$} when $w_0 \to w_\pm$.

\section{Examples: Maldacena-N\'u\~nez solutions}\label{secMN}

In this section we discuss the $\cN=1$ and $\cN=2$ Maldacena-N\'u\~nez solutions (MN1 and MN2)~\cite{Maldacena:2000mw}, which can be obtained as special cases of the BBBW solutions with enhanced global symmetry $SU(2)\times U(1)$.  For MN1 the $SU(2)$ is a flavor symmetry, and for MN2 it is an R-symmetry. 

\subsection{$\cN\!=\!1$ MN solution} \label{secMN1}

The MN1 solution is an important example since the dual SCFT has the lowest central charge amongst the BBBW theories.
It is obtained by setting $z=0$ and $\kappa=-1$, which gives
\begin{equation}
w_- = \frac{1}{18} = w_+ ~, \qqq    \ex^{2\nu} = \frac{1}{3}~. \label{MN1defs}
\end{equation}  
Since the endpoints $w_\pm$ of the $w$-interval are equal, it would naively appear that the metric is singular, but this is actually a coordinate artifact. 
We can reproduce the MN1 metric in the form given in section~5.4 of~\cite{Gauntlett:2004zh} (the angles there have been given tildes here) by first performing the following coordinate transformation:
\bea
y&=&  \sqrt{6w_+w_-} \cos \tilde\alpha~, \nn
w &=&\frac{1}{2}\left[ w_++ w_- + (w_+-w_-)  \cos \tilde\theta \right]~,  \nn
\chi&=&  \frac{\ex^{-2\nu}}{2-2g}\frac{(w_++w_-)^2}{w_+ -w_-}   \tilde\nu~,
\eea
and then setting $w_+ = w_-$. 
This leads to\footnote
{
Here $\tx_i$ are the coordinates of the Poincar\'e half-plane model of hyperbolic geometry, as opposed to $x_i$ used in the rest of the paper, which are coordinates on the unit disk.
}
\bea\label{M6onMN1}
\dd s^2(M_6) &=& \frac{1}{3} \Bigg[\frac{\dd \tx_1^2 + \dd \tx_2^2}{\tx_2^2}  +  \dd \tilde\alpha^2 
+ \frac{\sin^2\tilde\alpha}{ 3+\cos^2\tilde\alpha} \left(\dd\tilde\theta^2 + \sin^2 \tilde\theta \dd\tilde\nu^2 \right)  \nn
&& \qquad +\frac{\sin^2\tilde\alpha}{ 3+\cos^2\tilde\alpha } \left(\dd\psi - \cos \tilde\theta \dd\tilde\nu -\frac{\dd \tx_1}{\tx_2}  \right)^2  \Bigg]~,
\eea
as well as $\ex^{6\lambda} = 6w_+w_- ( 3 + \cos^2\tilde\alpha)$.
The angles have the following ranges
\be
0\le \tilde \alpha \le \pi~, \qqq  0\leq \tilde\tha\leq \pi~, \qqq 0\leq \tilde\nu\leq 2\pi~,\qqq 0\leq \psi\leq 4\pi~.
\ee
We can see that $\{ \tilde\tha,\tilde\nu\}$ describe an $S^2$, which combines with $\psi$ into a round $S^3$, which itself combines with $\tilde \alpha$ to give a squashed $S^4$.
The background thus has the symmetry $U(1)_R\times SU(2)_F$, where the R-symmetry corresponds to the Killing vector field $\del_\psi$.

In order to preserve this $U(1)_R$, a probe M5-brane must wrap $S^1_\psi$.
It then sits at a constant position in the remaining directions. The BPS condition $y=0$ fixes $\tilde \alpha=\pi/2$, but leaves the position $\{\tilde \theta_0,  \tilde\nu_0\}$ on the $S^2$ arbitrary. Note that since the $S^2$ has a round metric this preserves $U(1)_F$.
The worldvolume of the probe M5-branes is thus given by
\bea
\text{\bf M5:} \qquad  \text{AdS$_5$} \times S^1_\psi \times  \{\tilde \alpha =\tfrac{\pi}{2} \} \times\{\text{point} \in S^2  \}  \times  \{\text{point $\in\Sigma_g$}\} . 
\eea
We see that there is no special point on the $S^2$, which is in agreement with the fact that for $w_+=w_-$ the M5-branes {\bf M5$_\pm$} and {\bf M5$_0$} are indistinguishable.
The multiplicity of the M5-brane is also everywhere the same since the winding numbers are $m_\pm=1$.

A probe M2-brane ending on the $S^1_\psi$ wrapped by the M5-brane extends from $\tilde\alpha=\pi/2$ to either point $\tilde\alpha=\{0,\pi\}$ where the $S^1_\psi$ shrinks.
The area of the cups wrapped by such a pair of M2-branes corresponding to the conformal dimensions of dual BPS operators is 
\bea 
\Delta[\text{\bf M2} ]  =  \Delta[ \widetilde{ \mathrm{ \bf M2}}]  =  \frac{3}4 N ~.
\eea

\subsection{$\cN\!=\!2$ MN solution}\label{secMN2}

The MN2 solutions are obtained by setting $z=-1$ (that is $p=0$) and $\kappa=-1$, which gives
\begin{equation}\label{wpmMN2}
w_- = \frac{1}{24}~, \qqq w_+= \frac{1}{12}~, \qqq \ex^{2\nu} = \frac{1}{2}~.
\end{equation}  
To obtain the MN2 metric in a form that makes the $SU(2)\times U(1)$ R-symmetry manifest, we make the coordinate transformation\footnote
{
The angles with tildes are those that appear in the MN2 metric as given in (2.7) of~\cite{Gaiotto:2009gz}.
}
\bea
y &=& \frac{1}{4\sqrt{3}} \cos\tilde\theta \cos \tilde \psi ~,   \qqq\qqq \,   \phi_+ =  \tilde \phi~, \nn
w &=& \frac1{24} \frac{\sin^2\tilde\theta + 2 \cos^2 \ttha \sin^2 \tilde \psi}{ \sin^2\tilde\theta + \cos^2 \ttha \sin^2 \tilde \psi} ~,   \qqq   \phi_- = - \tilde \chi~, 
\eea
with $\tilde \theta \in [0,\pi/2]$ and $\tilde \psi \in [0,\pi]$.
The metric then becomes 
\bea
\dd s^2 (M_6) &=& 2 \frac{ \dd x_1^2 + \dd x_2^2 }{(1-x_1^2 - x_2^2)^2} + \frac12 \dd \tilde \theta^2 + \frac{\cos^2\tilde \theta}{2(1+\cos^2\tilde \theta)} \left(\dd\tilde\psi^2 + \sin^2\tilde \psi \dd\tilde \phi^2 \right) \nn 
&&+ \frac{\sin^2 \tilde\theta}{1+\cos^2\tilde \theta} \left[ \dd\tilde \chi + (2g-2)V \right]^2~,
\eea
and the warp factor is given by $24\ex^{6\lambda} = 1+ \cos^2\tilde\theta$.
The R-symmetry corresponds to the isometries of the $S^2$ parametrized by $\{\tilde \psi, \tilde \phi\}$ and of the $S^1$ parametrized by $\tilde \chi$.

The BPS condition $y=0$ for a probe M5-brane leads to $\tilde \psi=\pi/2$ and leaves $\tilde \theta_0$ as a modulus. The M5-brane wraps a closed curve on the torus $S^1_+\times S^1_-$ of slope $m_+/m_- = w_+/w_-=2$:
\bea
\text{\bf M5$(\tilde \theta_0,  \chi_0 )$:} \qquad  \text{AdS$_5$} \times S^1_R\big(\tfrac{m_+}{m_-}=2\big) \times  \{\tilde \psi =\pi/2\}\times\{\tilde \theta_0,  \chi_0  \}  \times  \{\text{point $\in\Sigma_g$}\} .  
\eea
At a generic point $\{\tilde \theta_0, \chi_0  \}$ on the moduli space, the M5-brane only preserves the $U(1)$ R-symmetry.
When it sits at the extremities of $\tilde \theta$ it preserves some enhanced global symmetry.

At the extremity $\tilde \theta_0 = \pi/2$, that is at $w_-$, the $S^2$ parametrized by $\{\tilde \psi,\tilde \phi\}$ shrinks. This means that {\bf M5$_-$} wraps $S^1_{\tchi}$ and preserves the full $SU(2)_R\times U(1)_R$. This is the probe M5-brane studied in the $\cN=2$ setting~\cite{Gaiotto:2009gz}. 
On the other hand, at $\tilde \theta =0$, that is at $w_+$, the circle $S^1_{\tchi}$ shrinks and {\bf M5$_+$} wraps $S^1_{\tilde \phi}$, thus preserving $U(1)_R\times U(1)_F$. 

Now recall that a single {\bf M5$_0$} wrapped on the closed curve around the torus $S^1_+\times S^1_-$ with the winding numbers $m_+=2$ and $m_-=1$ becomes a number $m_\pm$ of coincident {\bf M5$_\pm$} wrapped on $S^1_\pm$ at the endpoints $w_\pm$.
In reverse, we can say that a single {\bf M5$_-$} can move away from $w_-$ and explore its moduli space, where it breaks the supersymmetry to $\cN=1$. 
However, an {\bf M5$_+$} alone is stuck at $w_+$; it needs to pair-bond in order to take off and explore the moduli space, where it fuses into a unique M5-brane.

\begin{figure}[t]
\centering
\includegraphics[width=6.5in]{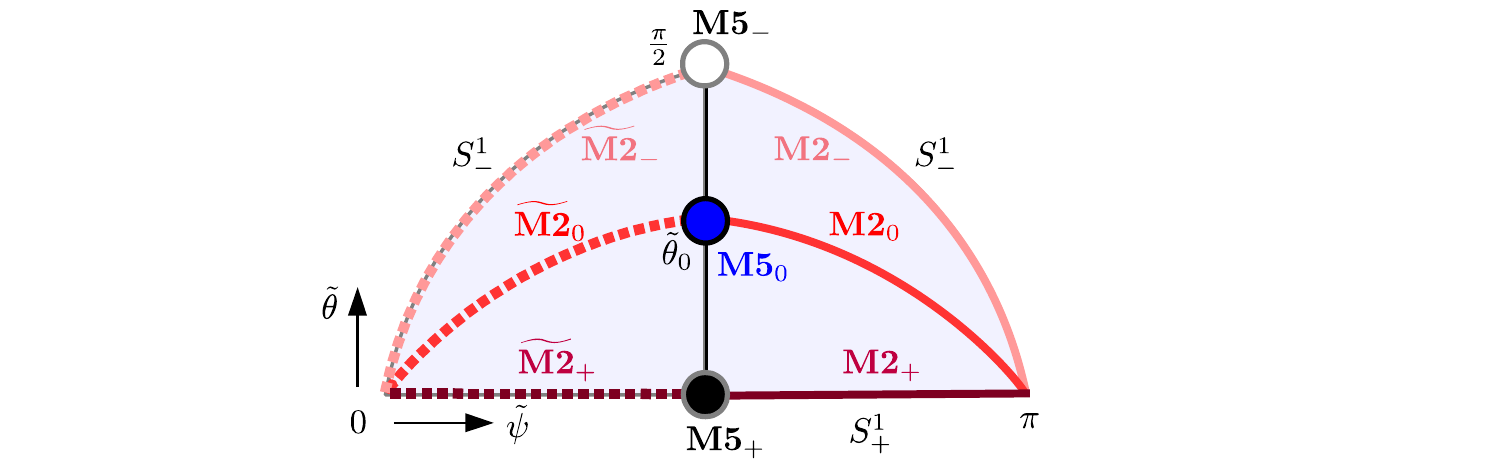}
\caption{Region parametrized by $\{\tilde \psi, \tilde \theta\}$ in the $\cN=2$ Maldacena-N\'u\~nez solution. The M5-branes (dots) are positioned along the axis $\tilde \psi = \pi/2$: {\bf M5$_\pm$} at $\tilde \theta = \{0,\pi/2\}$ and {\bf M5$_0$} at $\tilde \theta_0$. There is a pair of M2-branes (colored lines) ending on each M5-brane and extending towards $\tilde \theta=0$ with $\tilde \psi= \{0,\pi\}$.}
\label{MN2M5M2}
\end{figure}

A probe M2-brane ending on an M5-brane sitting at $\tilde\theta_0$ extends along the curve
\bea
\sin \tilde \psi = \frac{\cos \tilde \theta_0 }{\sin^2\tilde \theta_0}\frac{\sin^2 \tilde\theta }{  \cos \tilde\theta}~.
\eea
We show a few examples in figure~\ref{MN2M5M2}.
The conformal dimensions of the dual BPS operators is given by~\eq{dimAprobe} and~\eq{Deltaw0}:
\bea
\Delta[\text{\bf M2}_0 ]  &=&   \Delta[ \widetilde{ \mathrm{ \bf M2}}_0]  = N~,\nn
\Delta[\text{\bf M2}_- ]  &=&  \Delta[ \widetilde{ \mathrm{ \bf M2}}_-]  = N ~, \qqq
\Delta[\text{\bf M2}_+ ]  =  \Delta[ \widetilde{ \mathrm{ \bf M2}}_+]  = \frac N2~.
\eea
Note that these dimensions are consistent with the fact that it takes two {\bf M2$_+$} to make a single {\bf M2$_0$} or {\bf M2$_-$}.
The conformal dimension of {\bf M2$_-$} agrees with the result in~\cite{Gaiotto:2009gz}.

\section{Generalized quiver gauge theories and punctures}\label{secCFT}

In the previous section we have studied probe M5-branes in AdS$_5$ solutions of M-theory corresponding to the near horizon of $N$ M5-branes wrapping a Riemann surface $\Sigma_g$.
We have found various supersymmetric embeddings such that the probe M5-brane extends along all of AdS$_5$ and sits at a point of $\Sigma_g$.  
In this section we interpret this extra M5-brane in terms of the dual SCFTs. These gauge theories can be constructed from generalized quivers with the same topology as the Riemann surface $\Sigma_g$. The extra M5-brane can thus be viewed as a puncture (or a node) on the generalized quiver.

We first review this construction for $\cN=2$ SCFTs, and then describe the $\cN=1$ generalized quivers dual to the BBBW solutions on which we focused.
We stress that whereas the gravity solutions allow all values of $p$ and $q$, only the field theories with $p,q\ge0$ have been constructed.

The reader familiar with generalized quivers can safely skip subsections~\ref{n2scfts} and~\ref{n1scfts}, and go directly to subsection~\ref{puncturesBPS}, where we match the volumes of the M2-branes studied above to the conformal dimensions of BPS operators associated to a simple puncture in $\cN=1$ generalized quivers.

\subsection{Review of $\cN\!=\!2$ generalized quivers}\label{n2scfts}

The $\cN=2$ SCFTs arising from the low-energy limit of $N$ M5-branes that are wrapping a Riemann surface with punctures can be described in terms of generalized quivers \cite{Gaiotto:2009we}.
The building blocks for these theories are isolated $\cN=2$ SCFTs that described the low-energy dynamics of $N$ M5-branes wrapping a thrice-punctured sphere (trinion).   
Each puncture is associated with a global symmetry that is a subgroup of $SU(N)$.  

\begin{figure}[t]
\centering
\includegraphics[width=6.5in]{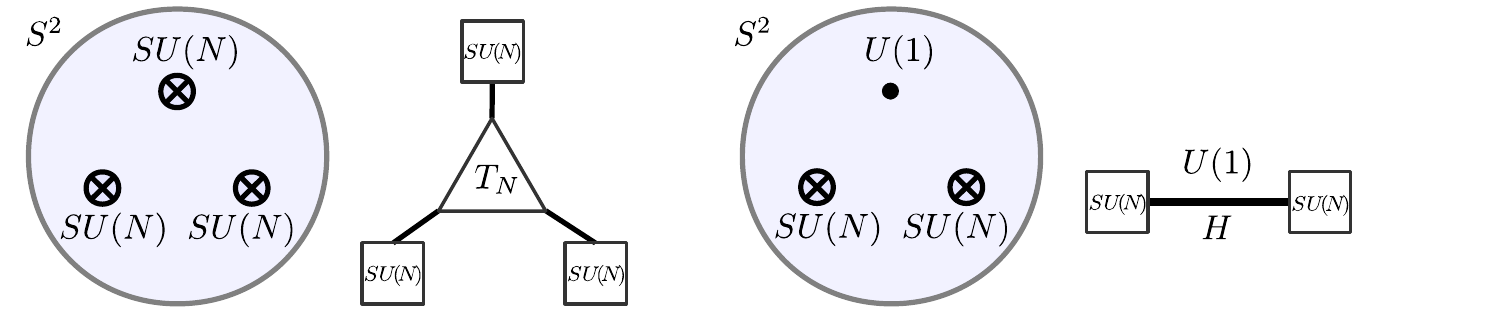}
\caption{\emph{Left}: sphere with three maximal punctures, corresponding to the theory $T_N$ with three $SU(N)$ global symmetries. \emph{Right}: sphere with two maximal and one minimal punctures, corresponding to a bifundamental hypermultiplet $H$ with a $U(1)$ global symmetry.}
\label{trinion}
\end{figure}

A \emph{maximal} puncture has an $SU(N)$ global symmetry. Locally, it can be understood as the branching of $N$ M5-branes wrapping a circle that shrinks at infinity.  
The origin of the $SU(N)$ global symmetry becomes clear by thinking about this configuration as a stack of $N$ semi-infinite D4-branes in type IIA string theory. 
The $T_N$ theory is the isolated $\cN=2$ SCFT that arises from $N$ M5-branes on a sphere with three maximal punctures (see figure~\ref{trinion}, left).
For each maximal puncture there is a Higgs branch operator $\mu_\alpha$ that transforms in the adjoint of the associated $SU(N)$ and has conformal dimension~$2$.   
These operators satisfy the chiral ring relation $\mu_1^2 =\mu^2_2=\mu^2_3$, where the square implies a trace over the adjoint indices;  they are neutral under $U(1)_R$ and carry charge $1$ under the Cartan $U(1)$ of $SU(2)_R$.  The Coulomb branch is parametrized by some operators $u_k^{(1)}$, with $k = 3, ..., N$ and $i = 1, ..., (k - 2)$, which have scaling dimension $\Delta[u^{(i)}_k] = k$. Finally, there are trifundamental operators in the $(\bf N, N, N)$ and $(\bf \bar{N}, \bar{N}, \bar{N})$ of $SU(N)^3$ which have dimension $(N - 1)$.

Punctures associated with smaller subgroups of $SU(N)$ can be obtained by higgsing maximal punctures.  
A \emph{simple} puncture has a $U(1)$ global symmetry.  It corresponds to the branching of a single M5-brane, which reduces to an NS5-brane in type IIA string theory. 
Wrapping $N$ M5-branes on a sphere with two maximal and one simple punctures gives rise to an  hypermultiplet $H$ in the bifundamental of $SU(N)\times SU(N)$, which comes with a $U(1)$ baryonic flavor symmetry (see figure \ref{trinion}, right). In $\cN=1$ language, this hypermultiplet can be described as a pair of chiral multiplets in conjugate representations, $H= (Q, \widetilde{Q})$, which forms an $SU(2)_R$ doublet $(Q, \widetilde{Q}^\dagger)$; these chiral fields are neutral under $U(1)_R$, carry charges $(1/2,1/2)$ under the Cartan of $SU(2)_R$, and $(1,-1)$ under the $U(1)$ flavor symmetry.  The two adjoint operators associated to the maximal punctures are $\mu_1 = (Q\widetilde{Q})$ and $\mu_2= (\widetilde{Q} Q)$ (one contracts the indices of the second $SU(N)$ for $\mu_1$, and the indices of the first $SU(N)$ for $\mu_2$).  The chiral ring relation $\mu_1^2 =\mu_2^2$ is automatic.   

$\cN=2$ generalized quivers are built from trinions by gluing punctures pairwise. This corresponds to gauging pairs of global symmetries and identifying them. 
The reduction of the worldvolume theory of the $N$ M5-branes on a cylinder joining two maximal punctures yields an $\cN=2$ vector multiplet $\cA$ in four dimensions.  In $\cN=1$ language, this gives an $\cN=1$ vector multiplet $A$ and a chiral superfield $\phi$ in the adjoint of the $SU(N)$ gauge group, $\cA = (A, \phi)$.  The chiral adjoint has charge $2$ under $U(1)_R$ symmetry and is neutral under $SU(2)_R$.  $\cN=2$ supersymmetry requires a superpotential term $W= \phi \mu$ for each gauged puncture, with fixed coupling.

\begin{figure}[tbh]
\centering
\includegraphics[width=6.5in]{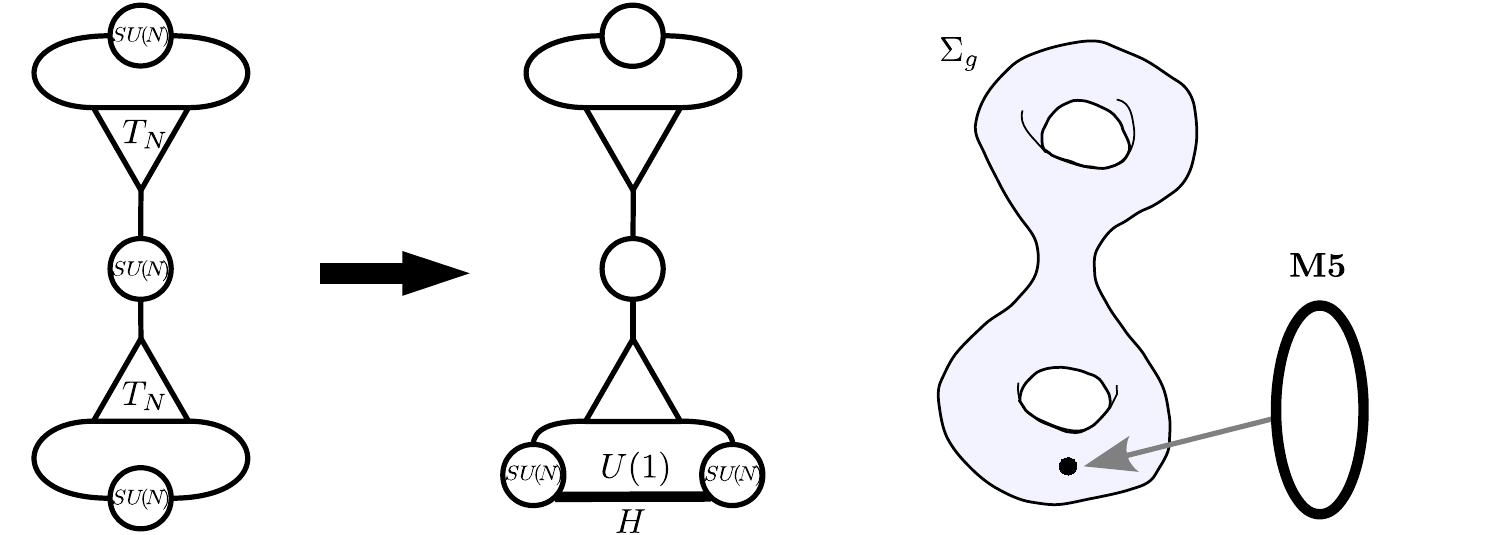}
\caption{\emph{Left}: A generalized quiver with $T_N$ building blocks (triangles) and $SU(N)$ gauge groups (circles). \emph{Middle}: insertion of a hypermultiplet $H$ in the bifundamental of $SU(N)\times SU(N)$.
\emph{Right}: the corresponding Riemann surface $\Sigma_g$ with an extra M5-brane intersecting it at a point (see~\cite{Gaiotto:2009gz}).}
\label{quiverGM}
\end{figure}

The generalized quiver corresponding to $N$ M5-branes wrapping a genus $g$ Riemann surface $\Sigma_g$ without punctures is obtained by fully gluing together $2(g-1)$ $T_N$ theories with $3(g-1)$ $\cN=2$ vector multiplets \cite{Gaiotto:2009we} (see the example in figure~\ref{quiverGM}, left).  The gravity duals are the MN2 solutions that we have discussed in section~\ref{secMN}.  
In contrast, when some of the punctures (which can be higgsed) are not glued, the resulting field theory has additional flavor symmetries. 
The gravity duals of such theories were described in~\cite{Gaiotto:2009gz} in terms of 
a general class of $\cN = 2$ solutions (LLM)~\cite{Lin:2004nb}.  
However, the addition of a single extra M5-brane, that is a simple puncture, to a Riemann surface without punctures can be analyzed in the probe approximation within the MN2 solutions. 
In the dual field theory, this corresponds to the insertion of a sphere with two maximal and one simple punctures (a hypermultiplet and a $U(1)$ flavor symmetry) into a generalized quiver (see figure~\ref{quiverGM}). 

\subsection{Review of $\cN\!=\!1$ generalized quivers}\label{n1scfts}

The $\cN=1$ SCFTs dual to the BBBW solutions (with $p,q\ge0$) are constructed in a similar way to the $\cN=2$ generalized quivers.  The novelty is that maximal punctures can be glued with $\cN=1$ vector multiplets as well as of $\cN=2$ vector multiplets~\cite{Bah:2011je,Bah:2012dg}.\footnote
{
Gluing trinions with $\cN = 1$ vector multiplets can be described in field theory, but an understanding from M5-brane point of view is still lacking.
}
Since the BBBW geometries have an $U(1)_R\times U(1)_F$ isometry, the gluing must be such that the resulting SCFT preserves two $U(1)$ global symmetries.
The $\cN=1$ SCFTs dual to the MN1 solutions can be obtained by mass deformation of the $\cN=2$ SCFTs dual to the MN2 solutions~\cite{Benini:2009mz}.

We first describe how to glue two trinions by a pair of maximal punctures.
Since we are constructing $\cN=1$ theories, we think about the $\cN=2$ theory of a trinion from an $\cN=1$ point of view.
We denote the generator of the Cartan of $SU(2)_R$ by $I_3$, and the generator of $U(1)_R$ by $R_{\cN=2}$.
Each trinion theory can then be viewed as a $\cN=1$ theory with an R-symmetry generated by $R_0$ and a $U(1)$ flavor symmetry generated by $F$:
\begin{equation}
R_0 = \frac{1}{2} R_{\cN=2} +I_3~, \qqq J = \frac{1}{2}R_{\cN=2}-I_3~,
\end{equation} 
The $\cN=1$ superconformal R-symmetry is generated by
\begin{equation}\label{RR0J}
R= R_0 - \frac{1}{3} J= \frac{1}{3} R_{\cN=2}+\frac{4}{3} I_3 ~ .  
\end{equation}  
When two trinions with flavor symmetries $J_1$ and $J_2$ are glued by a pair of maximal punctures with $\cN=1$ and $\cN=2$ vector multiplets, there are constraints from chiral anomalies.  
The contribution of a trinion theory to the anomaly is the same as that of $N$ fundamental hypermultiplets~\cite{Gaiotto:2009we}, which gives $-N/2$ for $R_0$ and $-N/2$ for $J$.  
Observe that the gluing always preserves $R_0$ since the anomaly contributions from the trinions are cancelled by the gaugino contribution in the vector multiplet.  On the other hand $J_1$ and $J_2$ are both anomalous, but $J_1-J_2$ is preserved.  The gluing is completed by adding the superpotential term $W= \mu_1 \mu_2$.  

When we glue trinions with $\cN=2$ vector multiplets $\cA = (A, \phi)$, we also need to consider the anomaly contributions of chiral adjoint $\phi$, which comes with a $U(1)$ flavor symmetry $F$.
This $\cN=2$ gluing preserves an additional $U(1)$ symmetry generated by $J_1+J_2 +F$.
However, $J_1-J_2$ is broken by the $\cN=2$ superpotential term $W= \phi (\mu_1 + \mu_2)$, which is necessary in order for the $\cN=2$ vector multiplet not to decouple in the IR~\cite{Bah:2012dg,Bah:2013aha}.

$\cN=1$ generalized quivers can be obtained by gluing trinions with a judicious combination of $\cN=1$ and $\cN=2$ vector multiplets on maximal punctures, in such a way that they preserve an R-symmetry $R_0$ and a flavor symmetry $\mathcal{F}$. The flavor symmetry acts on a trinion either as $J$ or as $-J$.
This can be achieved by attributing a sign $\sigma_i = \pm 1$ to each trinion, and gluing pairs of trinions with the identical or opposite signs by $\cN=2$ or $\cN=1$ vector multiplets, respectively.  The flavor $U(1)$ is written as
\begin{equation}
\cF = \sum_{i} \sigma_i J_i + \frac{1}{2} \sum_{\langle i,j\rangle} (\sigma_i + \sigma_j) F_{i_a j_b}~,
\end{equation}  
where $F_{i_a j_b}$ is the flavor $U(1)$ for the chiral adjoint in the vector multiplet connecting the $a$th maximal puncture in the $i$th trinion with the $b$th maximal puncture in the $j$th trinion.  The first sum is over the trinions while the second is over the vector multiplets.  The $\cN=1$ superconformal R-symmetry of the field theory can be written as
\begin{equation}\label{RR0eF}
R = R_0 + \epsilon \mathcal{F}
\end{equation} 
where the parameter $\epsilon$ can be determined by $a$-maximization (see~\cite{Bah:2011je,Bah:2012dg,Bah:2013aha} for more detail).

\begin{figure}[t]
\centering
\includegraphics[width=3in]{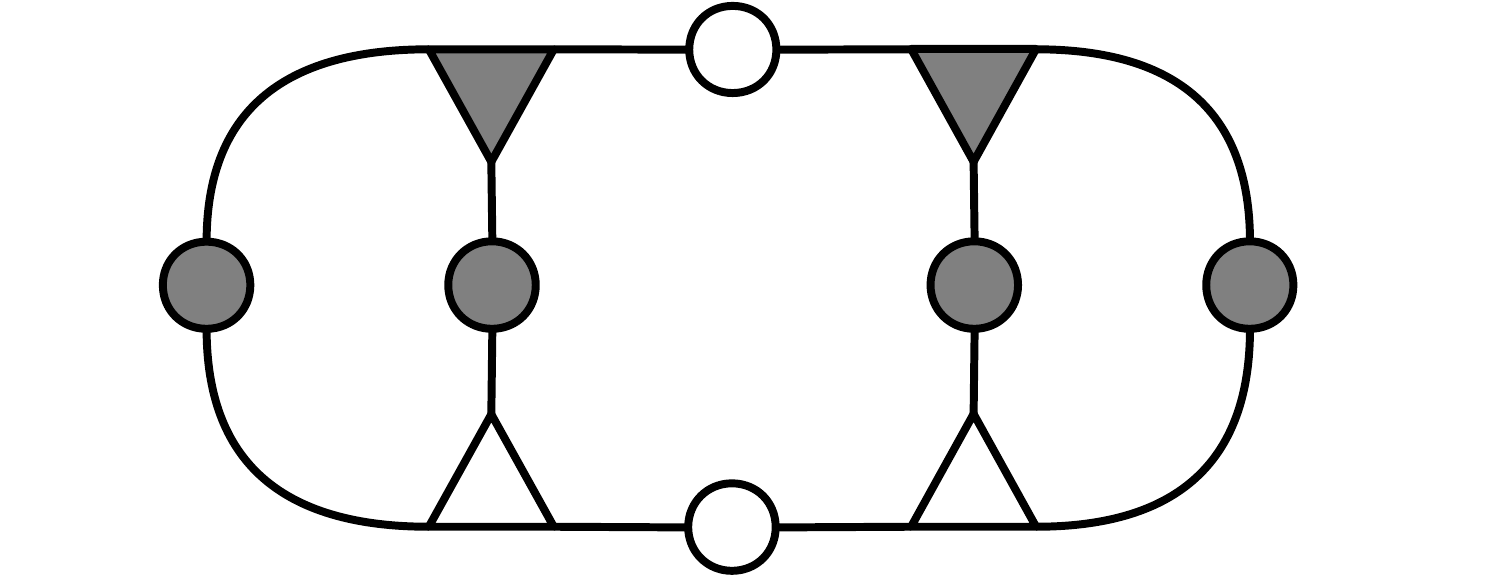}
\caption{Example of a colored generalized quiver with $p$ black $T_N^+$and $q$ white $T_N^-$ theories (here for $p=q=2$). There are $\cN=2$ vector multiplets (white) between pairs of theories with the same color, and $\cN=1$ vector multiplets (black) between theories of opposite colors.}
\label{BBBWquiver}
\end{figure}

The $\cN=1$ SCFTs dual to the BBBW solutions with a Riemann surface $\Sigma_g$ are obtained by gluing all the maximal punctures of a number $2(g-1)$ of $T_N^{\sigma_i}$ theories, which split into $p$ black $T_N^+$ and $q$ white $T_N^-$~\cite{Bah:2012dg} (see the example in figure~\ref{BBBWquiver}).
We see that this construction only makes sense for $p,q \geq0$.
Note that when $p=0$ or $q=0$, we get an $\cN=2$ theory as in the previous subsection. The theories with $p=q = g-1$ are dual to the MN1 solutions.  

After $a$-maximization, the parameter $\epsilon$ and the central charge of the theories depend only on $N$, $p$ and $q$. In the large $N$ limit, this gives
\begin{equation}
\epsilon = \frac{1 - \sqrt{1+ 3z^2}}{3z}  ~.  \label{epscft}
\end{equation} 
Since $-1\le z\le 1$ from $p,q\ge0$,\footnote
{
Note that in this section we are not restricting the sign of $z$ as in previous sections, but switching the sign of $z$ just exchanges black and white $T_N$ theories, as we will see momentarily.
}  this parameter $\epsilon$ satisfies the following bound (also valid for finite $N$):
\begin{equation}\label{boundeps}
-\frac{1}{3} \leq \epsilon \leq \frac{1}{3}~.  
\end{equation}  
This bound guarantees that the conformal dimensions of gauge-invariant chiral operators do not violate the unitarity bound~\cite{Bah:2012dg}. 
The lower bound is saturated by the $\cN=2$ theory with only black $T_N$ theories ($q=0$ and $z=1$) and the upper bound is saturated by the $\cN=2$ theory with only white $T_N$ theories ($p=0$ and $z=-1$).

In the BBBW theories, $R_0$ and $\mathcal{F}$ are expressed in terms of two generators $R_\pm$ that are dual to phase rotations ${\partial}/{\partial \phi_\pm}$ on the fibers of the line bundles $\cL_p$ and $\cL_q$: 
\begin{equation}\label{R0andF}
R_0 = \frac{1}2 \left(R_+ + R_- \right), \qqq   \mathcal{F} = \frac{1}{2} \left(R_+ -R_-\right)~.
\end{equation}
The superconformal R-symmetry~\eq{RR0eF} then takes the form
\begin{equation}  \label{U1R}
R = \frac{1+\epsilon}2 R_+  +   \frac{1-\epsilon}2  R_-  ~.
\end{equation}

\subsection{Punctures and BPS operators}\label{puncturesBPS}

We just reviewed how the $\cN=1$ BBBW field theories can be constructed by fully gluing $T_N$ theories of two types, $p$ of the type $T_N^+$ and $q$ of the type $T_N^-$.
This give rise to an $\cN=2$ vector multiplet for each pair of glued theories with identical signs, and to an $\cN=1$ vector multiplet for each pair with opposite signs.

\begin{figure}[tbh]
\centering
\includegraphics[width=6.5in]{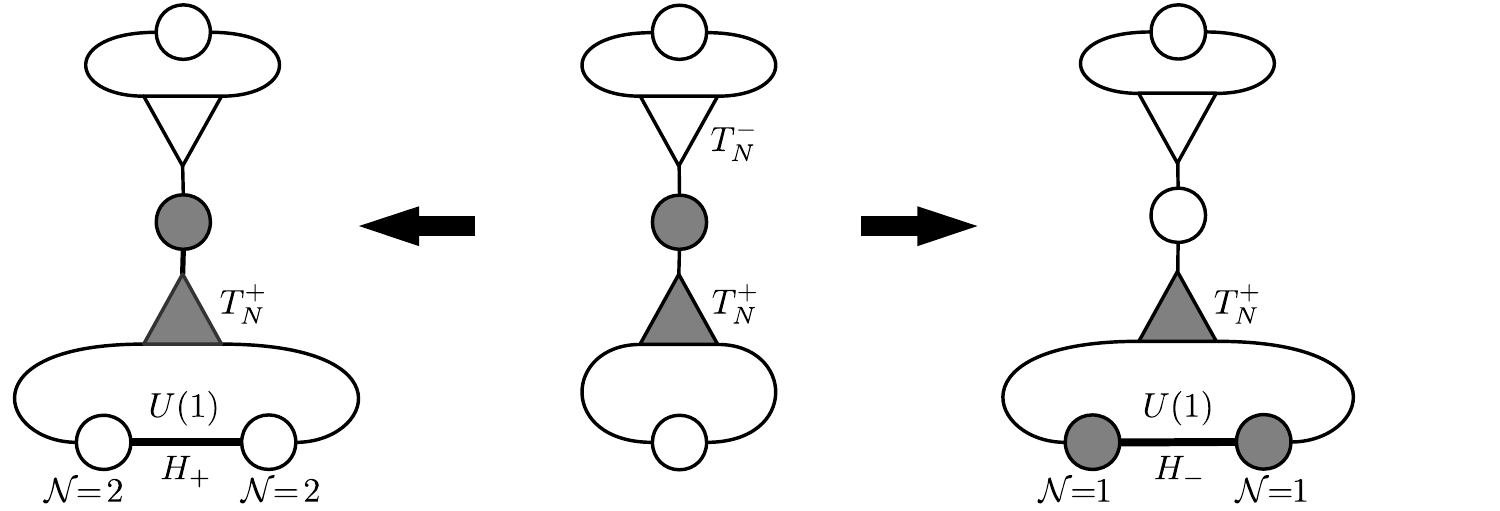}
\caption{\emph{Left}: insertion of a hypermultiplet $H_+$ next to a $T_{N}^+$ theory, with $\cN=2$ vector multiplets. \emph{Right}:~insertion of a hypermultiplet $H_-$ next to a $T_{N}^+$ theory, with $\cN=1$ vector multiplets.}
\label{quiverpuncture}
\end{figure}

Simple punctures can be created by inserting a hypermultiplet coming from a sphere with two maximal $SU(N)$ punctures and one $U(1)$ puncture~\cite{Bah:2011je,Bah:2013aha}.  There are also two types of hypermultiplets, denoted by $H_\pm= (Q_\pm, \widetilde{Q}_\pm)$. We show two examples in figure~\ref{quiverpuncture}.
It was observed in~\cite{Bah:2011je} that at large $N$ the $a$-maximization is dominated by the $T_N$ theories when the number of hypermultiplets is much smaller than $N$.  This is due to the fact that the $T_N$ theories contribute as $N^3$ to the central charge, while the hypermultiplets contribute as $N^2$, which is also true for BBBW theories with simple punctures.  We conclude that at large $N$, adding a small number of simple punctures to a  BBBW quiver doesn't change the value of the parameter~$\epsilon$~\eqref{epsgr}.  This is the field theory analogue of studying extra M5-branes in the probe approximation, as we did in previous sections.  
In terms of the parameters $w_\pm$~\eq{acpart} appearing in the BBBW solutions,
the parameter $\epsilon$ in~\eq{epscft} takes the form:
\begin{equation}
\epsilon  = \frac{w_+ -w_-} {w_+ + w_-}~.  \label{epsgr}
\end{equation}
Notice that the superconformal R-symmetry~\eq{U1R} then takes the form
\begin{equation} 
R = \frac{w_+} {w_+ + w_-} R_+  +  \frac{w_-} {w_+ + w_-}  R_-  ~,
\end{equation}
which matches nicely the expression~\eq{Killingvf} for the dual Killing vector field $\del_\psi$.

The chiral fields in hypermultiplets $H_\pm= (Q_\pm, \widetilde{Q}_\pm)$ have charges $(R_0, \cF)=(1/2, \mp1/2)$.  Using~\eq{R0andF} and~\eq{U1R}, we then find the R-charges
\begin{equation}\label{RchargeQpm}
R [Q_\pm] = R [\tQ_\pm] = \frac{1\mp \epsilon}{2}  = \frac{w_\mp}{w_+ +w_-}~.
\end{equation}  
The conformal dimensions of gauge-invariant chiral operators are related to their R-charges via $\Delta = 3R/2$.
We can construct gauge-invariant chiral operators by taking the determinants of $Q_\pm$ and $\tQ_\pm$, which simply multiplies the conformal dimensions by a factor of $N$:
\begin{equation}
\Delta [\det Q_\pm] = \Delta [\det \tQ_\pm] = \frac{3N}{2}   \frac{w_\mp}{w_+ +w_-}~.
\end{equation}
This agrees exactly with the volumes~\eq{dimAprobe} of the cups wrapped by the probe M2-branes ending on the M5-branes located at $w_\pm$.
We thus conclude that the two types of M2-branes {\bf M2$_\pm$} and $\widetilde{ \mathrm{ \bf M2}}_\pm$ (extending towards $y_0$ or $-y_0$) are dual to the two BPS operators $\det Q_\pm$ and $\det \tQ_\pm$.
This provides evidence that the simple punctures $H_\pm$ in the quivers correspond to the probe M5-branes {\bf M5$_\pm$} in the AdS$_5$ solutions.  
Note also that the $U(1)$ flavor symmetry of $H_\pm$ corresponds to the $U(1)$ gauge field in AdS$_5$ that arises from the Kaluza-Klein reduction on $S^1_\pm$ of the self-dual two-form potential on the worldvolume of {\bf M5$_\pm$}~\cite{Gaiotto:2009gz}.

The $\cN=1$ generalized quiver theories discussed here always have a $U(1)_F$ symmetry and so cannot describe punctures corresponding to probe M5-branes {\bf M5$_0$} at $w_0\neq w_\pm$. This hints at a larger class of punctures that only preserve the $U(1)$ R-symmetry (also recently pointed out in \cite{Xie:2013gma}).  The field theory interpretation of the probe M2-branes ending on the M5-branes located at $w_0\neq w_\pm$ is therefore unclear to us.

\subsection{$\cN\!=\!1$ punctures and MN theories}

The punctures associated to {\bf M5$_\pm$} locally preserve $\cN=2$ supersymmetry and have been studied in field theory by \cite{Beem:2012yn,Xie:2013gma,Bah:2013aha}.
By contrast, the puncture associated to {\bf M5$_0$} only preserve $\cN=1$ supersymmetry and have no known counterparts in field theory.
In this subsection, we briefly discuss what becomes of them in the field theories dual to the $\cN=1$ and $\cN=2$ MN solutions.

First, recall that {\bf M5$_0$} probes, and hence the associated punctures, only exist for BBBW theories where $\epsilon$~\eq{epsgr} is rational. Note that this is the same condition as the one for a rational central charge discussed in subsection~\ref{RationalCC}.
This is of course satisfied by all the MN theories, for which $\epsilon =0$ (MN1) or $\epsilon = \pm 1/3$ (MN2).  Between these two cases, there are infinitely  many rational BBBW theories, but the field theory requirement $p,q\ge0$ makes them rather sparse.  
Up to genus $25$ there are only 8 rational SCFTs, the lowest one being at $g=12$ with $(p,q)=(15,7)$.
We can write an infinity family in terms of the integers $n_1$ and $n_2$ appearing in~\eq{zSol} and a positive integer $n>0$:
\bea
g = 1 + (3n_1^2 - n_2^2)n ~, \qqq  \begin{array}{lr}
   p= (n_1+n_2)(3n_1-n_2) n~, \\
    q= (n_1-n_2)(3n_1+n_2) n ~,
  \end{array} 
\eea
with the condition $n_1>n_2\ge 1$. We now turn to the MN solutions.

For the MN1 solutions, we saw in subsection~\ref{secMN1} that {\bf M5$_\pm$} and {\bf M5$_0$} are all identical, so that the position on the $S^2$ moduli space does not matter.  In the dual field theory, this corresponds to the fact that the R-charges~\eq{RchargeQpm} of $H_+ $ and $H_-$ agree, given that $\epsilon=0$. 
Moreover, given that all the chiral fields have R-charges~\eq{RchargeQpm}
equal to $1/2$, we can construct the following two gauge-invariant marginal operators:
\be
\cO_{4}^{\pm}= (Q_\pm \tQ_\pm)^2~.
\ee
It was shown in~\cite{Benini:2009mz} that each simple puncture provides two complex parameters, one for the position of the M5-brane on the Riemann surface $\Sigma_g$, and the other for the $\mathbb{CP}^1$-worth of directions in the fiber over $\Sigma_g$.
So we could naturally interpret the marginal operators $\cO_4^\pm$ as the movement of {\bf M5$_\pm$} on the $S^2$ moduli space.

For the MN2 solutions with $p=0$ we get $\epsilon = 1/3$, which gives the R-charges $R [Q_+] =R [\tQ_+] =  1/3 $ and $R [Q_-] = R [\tQ_-] =  2/3$.
In this case, we can only construct one gauge-invariant marginal operators:
\be
\cO_{6}^+ = (Q_+ \tQ_+)^3~.
\ee
However, in this case, the corresponding M5-brane is fractional and cannot move away from the pole unless it pairs up with a second M5-brane. Vexingly, it is the M5-brane at the other pole that can move alone on the $S^2$ moduli space. 
So we do not find a direct interpretation of this marginal operator.
A more complete analysis of the marginal operators as in~\cite{Benini:2009mz} might resolve this issue.

The field theories dual to the MN solutions are the only ones for which one of the winding numbers $m_\pm$ is equal to $1$, or in other words for which a single M5-brane can move away from a pole of the $S^2$ moduli space.
To see this, we note that the bound~\eq{boundeps} on $\epsilon$ implies the following bound:
\begin{equation}
\frac{1}{2} \leq\frac{m_+}{m_-} \leq 2~.
\end{equation}  
We see that if either one of $m_\pm$ is $1$, then the other one can only be $1$ (MN1) or $2$ (MN2).

That being said, recall that in the gravity side there is no reason to restrict to $p,q\ge0$ (that is $-1\le z\le 1$).
There is still a bound on $m_+/m_-$ in~\eq{wpmmpm}, which is only slightly larger (for $\kappa=-1$):
\bea
2 - \sqrt{3} \le \frac{m_+}{m_-} \le 2 + \sqrt3~.
\eea
This allows for one additional solution with a unit winding number, namely $m_+=3$, $m_-=1$ (given that we took $m_+\ge m_-$). 
This corresponds to $z=-4$ with $p=3(g-1)$, $q=-5(g-1)$, and gives
\bea
w_- = \frac1{27}~, \qqq w_+ = \frac1 9~, \qqq \ex^{2\nu} = \frac43~.
\eea
The dimensions of the BPS M2-branes are
\bea 
\Delta[\text{\bf M2}_+ ]  &=&   \frac{3N}{8} ~, \nn
\Delta[\text{\bf M2}_- ] &=&  \frac{9N}{8} = \Delta[\text{\bf M2}_0 ]  ~.
\eea
We could then imagine constructing an octic marginal operator from {\bf M2$_+$} in the putative dual SCFT.
Just like for the MN2 solutions, we see that this operator is associated with {\bf M5$_+$} at $w_+$ with winding number $m_+=3$.
It would be interesting to understand this special theory in more detail.

\section{Outlook}

We have studied supersymmetric probe M5-branes in the AdS$_5$ backgrounds of BBBW~\cite{Bah:2012dg}, and interpreted them as simple punctures in the dual $\cN=1$ SCFTs. 
This generalizes the probe analysis in~\cite{Gaiotto:2009gz} of the $\cN=2$ field theory punctures~\cite{Gaiotto:2009we} in terms of the $\cN=2$ MN and LLM solutions~\cite{Maldacena:2000mw,Lin:2004nb}.
For BBBW solutions with rational central charge, the M5-branes have a two-dimensional moduli space $\cM_2$, on which they generically only preserve $U(1)_R$.
At the poles of $\cM_2$ the global symmetry enhances to $U(1)_R\times U(1)_F$ or to $SU(2)_R\times U(1)_R$.

If we take a single M5-brane at a generic point of the moduli space and move it to a pole, we end up with multiple M5-branes.
The same phenomenon happens to fractional branes at orbifold singularities~\cite{Gimon:1996rq}\cite{Douglas:1996sw}.
We saw that there are indeed $\ZZ_m$ singularities are the poles of the moduli space.

Incidentally, $\ZZ_m$ singularities play a key role for the construction of general punctures in the $\cN=2$ setting~\cite{Gaiotto:2009gz}.
In particular, a maximal puncture arises from a parabolic
element in the Fuchsian group $\Gamma \subset SL(2,\RR)$ by which hyperbolic space gets quotiented to produce a Riemann surface with a cusp: $\Sigma_g = \HH^2/\Gamma$. 
Because of the $\cN=2$ twist, this operation also induces some rotation in the fiber, so that the singularity is locally $\CC^2/\ZZ_N$ (see~\cite{Benini:2009mz} for the case of MN).
For the $\cN=1$ BBBW solutions, it is expected that more general orbifold singularities can appear, for example $\CC^3/\ZZ_m\times \ZZ_n$.

In order to understand intermediate punctures, whose flavor symmetry can be any subgroup of $SU(N)$, we must consider more general $\cN=1$ solutions than the BBBW solutions.
For the $\cN=2$ case, the appropriate geometry is provided by the general AdS$_5$ solutions (LLM)~\cite{Lin:2004nb}.
These solutions are determined in terms of a solution to the three-dimensional Toda equation, and the additional M5-branes act as boundary conditions for this equation~\cite{Gaiotto:2009gz}.
It would be interesting to understand the corresponding story for the $\cN=1$ case by studying boundary conditions for the equations of~\cite{Bah:2013qya}, which describe generalizations of the BBBW solutions with $U(1)^2$ symmetry. 
The {\bf M5$_0$} should also be associated with boundary conditions in the system of~\cite{Gauntlett:2004zh}.

Our work can be compared to the study of BPS probe D-branes in AdS$_5\times X_5$ solutions of IIB supergravity~\cite{Ouyang:2003df,Kuperstein:2004hy, Arean:2004mm, Canoura:2005uz} (see~\cite{Franco:2013ana} for recent work in this direction).
In particular, D7-branes wrapping a topologically trivial three-cycle in $X_5$ produce additional nodes on the quiver of the dual SCFT, corresponding to flavor symmetries. 
Before the backreaction, the D7-branes wrap a holomorphic four-cycle in a conical Calabi-Yau threefold $C(X_5)$.

Similarly, from a UV point of view, the probe M5-branes in this paper should wrap holomorphic two-cycles in the Calabi-Yau threefold $\cL_p\oplus \cL_q\to \Sigma_g$ mentioned in the introduction. Note that these cycles must be topologically trivial in order to avoid tadpoles.
We conjecture that {\bf M5$_\pm$} arise from M5-branes wrapping the two-cycles dual to the holomorphic four-cycles $\cL_p\to \Sigma_g$ and $\cL_q\to \Sigma_g$.
The interpretation of {\bf M5$_0$} appears more challenging.

We observed that the $S^1$-fibration over the two-dimensional base $\cM_2$ appears to be a Seifert fiber space.
This would imply in particular that two M5-branes at different points on the base $\cM_2$ wrap circles that can be linked. There might then be some Hanany-Witten effect at play.
It would be interesting to investigate the pertinence of this perspective.

The duality group of the $\cN=2$ theories was predicted by Witten in \cite{Witten:1997sc} to be the mapping class group of $\Sig_g$ with marked points. The moduli space of theories with punctures associated to the {\bf M5$_0$} includes a two-dimensional space $\cM_2$ fibered over $\Sig_g$. The duality group for such theories is surely interesting.

The approach of this paper can also be applied to solutions of M-theory involving AdS$_4\times \HH^3$ or AdS$_3\times \HH^4$.
Our work in this direction will appear elsewhere~\cite{BGH2}.

\acknowledgments

It is a pleasure to thank Ken Intriligator, Juan Maldacena, Jaewon Song, and Brian Wecht.
We would also like to thank the Simons Center for Geometry and Physics for hospitality during part of this project.
IB is supported in part by the DOE grant DE-FG03-84ER-40168, ANR grant 08-JCJC-0001-0, and the ERC Starting Grants 240210-String-QCD-BH, and 259133-ObservableString.  IB is grateful for the hospitality and work space provided by the UCSD Physics Department.
The work of MG has been supported in part by the ERC grant 259133, and by the German Science Foundation (DFG) within the Research Training Group 1670 ``Mathematics Inspired by String Theory and QFT.''

\bibliographystyle{jhep} 

\providecommand{\href}[2]{#2}\begingroup\raggedright\endgroup

\end{document}